\documentclass[aps,preprint,showpacs,nofootinbib]{revtex4}
\usepackage{amsfonts}
\usepackage{amsmath}
\usepackage{amssymb}
\usepackage{relsize}
\usepackage{graphicx}
\usepackage{epsfig}
\usepackage{float}
\usepackage[caption = false]{subfig}
\usepackage{color}

\usepackage{multirow}
\newcommand{\ic}{\textbf{\textbf{i}}}
\begin{document}

\title{Angular Fock coefficients. Fixing the errors, and further development.}

\author{Evgeny Z. Liverts}
\affiliation{Racah Institute of Physics, The Hebrew University, Jerusalem 91904,
Israel}

\author{Nir Barnea}
\affiliation{Racah Institute of Physics, The Hebrew University, Jerusalem 91904,
Israel}

\begin{abstract}
The angular coefficients $\psi_{k,p}(\alpha,\theta)$ of the Fock expansion characterizing the S-state wave function of the two-electron atomic system,
are calculated in hyperspherical angular coordinates $\alpha$ and $\theta$.
To solve the problem the Fock recurrence relations separated into the independent individual equations associated with definite power $j$ of the nucleus charge $Z$, are applied.
The "pure" $j$-components of the angular Fock coefficients, orthogonal to of the hyperspherical harmonics $Y_{kl}$, are found for even values of $k$.
To this end, the specific coupling equation  is proposed and applied.
Effective techniques for solving the individual equations with simplest nonseparable and separable  right-hand sides are proposed.
Some mistakes/misprints made earlier in representations of  $\psi_{2,0}$, were noted and corrected.
All $j$-components of $\psi_{4,1}$ and the majority of components and subcomponents of $\psi_{3,0}$ are calculated and presented for the first time.
All calculations  were carried out with the help of the Wolfram \emph{Mathematica}.
\end{abstract}


\maketitle

\section{Introduction}\label{S0}
The helium isoelectronic sequence presenting a two-electron atomic system contains the main features of a many-body system with Coulomb interaction. As such, it can serve as a simple basis for testing new quantum theories.
The state of a three-body system when all the particles are in same space point is known as the triple coalescence point (TCP).
A long time ago, Bartlett et al. \cite{B35} showed that the $^1 S$ helium wavefunction, $\Psi$ could not be expanded near the TCP as an analytic series in the interparticle coordinates $r_1,~r_2$ and $r_{12}$. Later Bartlett \cite{B37} and Fock \cite{FOCK} proposed the following expansion containing logarithmic functions
\begin{equation}\label{1}
\Psi(r,\alpha,\theta)=\sum_{k=0}^{\infty}r^k\sum_{p=0}^{[k/2]}\psi_{k,p}(\alpha,\theta)(\ln r)^p,
\end{equation}
where
\begin{equation}\label{2}
\alpha=2\arctan \left(r_2/r_1\right),~~~\theta=\arccos\left[(r_1^2+r_2^2-r_{12}^2)/(2r_1r_2)\right]
\end{equation}
are the hyperspherical angles, and $r=\sqrt{r_1^2+r_2^2}$ is the hyperspherical radius.
The method applied by Fock \cite{FOCK} for investigating the$~^1S$ helium wave functions was generalized \cite{ES1, DEM} for arbitrary systems of charged particles and for states of any symmetry.
The Fock expansion (FE) was used for treating the two hydrogen-atom system as the basic one for all subsequent calculations in the theory of dispersion forces \cite{TUL}.
The work of Fock was extended by expansion $\psi_{2,0}$ into hyperspherical harmonics (HH) \cite{ES2, DAV}.
The Fock expansion was somewhat generalized \cite{PLUV} to be applicable to any $S$ state, and its first two terms were determined.
The first numerical solution of the equations for the Fock coefficients was presented in work \cite{FMS}.
The most comprehensive investigation on the methods of derivation and calculation of the angular Fock coefficients
was presented in the works of Abbott, Gottschalk and Maslen \cite{AM1,GAM2,GM3}.
Methods for simplifying the recurrence relations generated by the Fock expansion (\ref{1}) were used \cite{AM1} to determine the highest power logarithmic terms to sixth order.
The wavefunction for $S$ states was given to second order in $r$ as a single and double infinite sums \cite{AM1}.
The results of ref. \cite{GAM2} hint at the existence of closed form wavefunction for few-body system.
The closed form of the helium-like wavefunction including terms up to second order in $r$ for$~^1S$ states, and up to forth order for$~^3S$
states was derived in \cite{GM3}.

In this paper, we build on the work \cite{AM1}, and therefore we try to adhere to the terminology used in this article.
We correct some substantial errors/misprints made in the final formulas for $\psi_{2,0}$ which is the basis for derivation of the representations for $\psi_{k,p}$ with $k > 2$.
We apply a new techniques for calculating an angular Fock coefficients (AFC), and for reducing some of them to the form of the one-dimensional series with fast convergence.
This technique is close to that of used in \cite{GM3}.
We separate the AFC into the components associated with definite power of the nucleus charge, and present all components of $\psi_{4,1}$ and the majority of components of $\psi_{3,0}$ for the first time.

We extensively use all the tools of the Wolfram Mathematica program (www.wofram.com/). Its last version, Mathematica 10 will be referred as \emph{Mathematica}. To make sure of the correctness of the analytical results, all of them are subjected to the numerical verification.

\section{General approach}\label{S1}

The Schr\"{o}dinger equation for a system of two electrons, in the field of an infinitely massive nucleus, is
\begin{equation}\label{100}
\left(-\frac{1}{2}\Delta+\boldsymbol V\right)\Psi=E \Psi ,
\end{equation}
where $E$ is the energy, and $\boldsymbol V $ is the Coulomb interaction. For a system with nucleus charge $Z$, it is useful to define the dimensionless potential $V\equiv\boldsymbol V r$, or
\begin{equation}\label{105}
V=\frac{1}{\sqrt{1-\sin \alpha \cos \theta}}-Z\left[\csc(\alpha/2)+\sec(\alpha/2)\right].
\end{equation}
The first term of the rhs of Eq.(\ref{105}) presents the electron-electron interaction, whereas the second one stands for the electron-nucleus interaction.
The Laplacian is
\begin{equation*}
\Delta=\frac{1}{r^5}\frac{\partial}{\partial r}r^5\frac{\partial}{\partial r}-\frac{1}{r^2}\Lambda^2,
\end{equation*}
where the hyperspherical angular momentum (HAM) operator, projected on $S$ states, reads:
\begin{equation}\label{104}
\Lambda^2=-\frac{4}{\sin^2 \alpha}\left(\frac{\partial}{\partial\alpha}\sin^2\alpha\frac{\partial}{\partial\alpha}+\frac{1}{\sin\theta} \frac{\partial}{\partial\theta}\sin\theta \frac{\partial}{\partial\theta}\right).
\end{equation}
By substituting the Fock expansion (\ref{1}) into the Schr\"{o}dinger equation (\ref{100}), one obtains the Fock recurrence relation (FRR) \cite{AM1}
\begin{subequations}\label{103}
\begin{align}
\left[ \Lambda^2-k(k+4)\right]\psi_{k,p}=h_{k,p},~~~~~~~~~~~~~~~~~~~~~~~~~~~~~~~~~~~~~~~~~~~~~~~~~~~~~~~~~~~~~\label{103a}\\
h_{k,p}=2(k+2)(p+1)\psi_{k,p+1}+(p+1)(p+2)\psi_{k,p+2}-2 V \psi_{k-1,p}+2 E \psi_{k-2,p}.\label{103b}
\end{align}
\end{subequations}
Atomic units are used throughout the paper.
It is important to note that $\psi_{k,p}=0$ for $k<0$ or $p>[k/2]$ (see, e.g., \cite{AM1}).

We shall solve the FRR (\ref{103}) to find the angular Fock coefficients, $\psi_{k,p}$.
It is well-known that the FRR can be solved by expanding the AFC into the hyperspherical harmonics (HH) of the form (see, e.g., \cite{AM1})
\begin{equation}\label{106}
Y_{kl}(\alpha,\theta)=N_{kl}\sin^l\alpha~C_{k/2-l}^{(l+1)}(\cos\alpha)P_l(\cos\theta),~~~~~~~~~~k=0,2,4,...; l=0,1,2,...,k/2
\end{equation}
where the $C_n^\nu(x)$ and $P_l(z)$ are Gegenbauer and Legendre polynomials, respectively. The normalization constant is
\begin{equation}\label{107}
N_{kl}=2^ll!\sqrt{\frac{(2l+1)(k+2)(k/2-l)!}{2\pi^3(k/2+l+1)!}},
\end{equation}
so that
\begin{equation}\label{108}
\int Y_{kl}(\alpha,\theta)Y_{k'l'}(\alpha,\theta)d\Omega=\delta_{kk'}\delta_{ll'},
\end{equation}
where $\delta_{mn}$ is the Kronecker delta, and the appropriate volume element is
\begin{equation}\label{109}
d\Omega=\pi^2 \sin^2\alpha~d\alpha \sin\theta d\theta.~~~~~~~~~~~~~~~~~\alpha\in [0,\pi],~\theta\in [0,\pi]
\end{equation}
The HH (\ref{106}) are the eigenfunctions of the operator $\Lambda^2$,  with eigenvalues given by $k(k+4)$.
They form a complete set of basis functions in $\left\{\alpha,\theta\right\}$.

Notice that any general function can be expanded into the hyperspherical harmonics
\begin{equation}\label{110}
\mathcal{F}(\alpha,\theta)=\sum_{nl}\mathcal{F}_{nl}Y_{nl}(\alpha,\theta),
\end{equation}
where the expansion coefficients are
\begin{equation}\label{111}
\mathcal{F}_{nl}=\int\mathcal{F}(\alpha,\theta) Y_{nl}(\alpha,\theta)d\Omega .
\end{equation}
Two important properties of the AFC must be emphasized:

\textbf{P1}. Any AFC, $\psi_{k,p}$ can be separated into the independent parts (components)
\begin{equation}\label{112}
\psi_{k,p}(\alpha,\theta)=\sum_{j=p}^{k-p} \psi_{k,p}^{(j)}(\alpha,\theta) Z^j
\end{equation}
 associated with definite power of $Z$, according to separation of the rhs (\ref{103b})
\begin{equation}\label{113}
h_{k,p}(\alpha,\theta)=\sum_{j=p}^{k-p} h_{k,p}^{(j)}(\alpha,\theta) Z^j
\end{equation}
of the FRR (\ref{103a}).
Hence, each of the FRR (\ref{103}) can be separated into the individual equations (IFRR) for each component:
\begin{equation}\label{114}
\left[ \Lambda^2-k(k+4)\right]\psi_{k,p}^{(j)}(\alpha,\theta)=h_{k,p}^{(j)}(\alpha,\theta).
\end{equation}
\textbf{P2}. Any component of the angular Fock coefficient, presenting a function of $\alpha$ and $\theta$, must be finite at each point of the two-dimensional angular space described by the hyperspherical angles $\alpha\in[0,\pi]$ and $\theta\in[0,\pi]$.

It is well-known that the general solution of the inhomogeneous equation can be expressed as the sum of the general solution of the associated homogeneous (complementary) equation, and the particular solution of the considered  inhomogeneous equation.
Note that the linear combination of HH, $Y_{kl}$ presents the general solution of homogeneous equation associated with inhomogeneous Eq.(\ref{114}).
It must be emphasized that the HH are defined by Eq.(\ref{106}) only for even values of $k$.
However, it may be verified that: a) the functions $Y_{kl}$ defined by Eq.(\ref{106}), but with odd values of $k$, also present the solutions of the homogeneous equation for Eq.(\ref{114}); b) the point $\alpha=\pi$ presents a singular one for $Y_{kl}(\alpha,\theta)$ with \textbf{odd} $k$.
It should be emphasized that this point presents a pole of the order $(l+1)$, and hence, no linear combination of $Y_{kl}$  with different $l$ (for given odd $k$) cannot eliminate this singularity.
The important conclusion is: for \textbf{odd} $k$ the finite particular solution of Eq.(\ref{114}) presents the physical solution we are looking for.
The exception is the case of the particular solution, which is singular at the point $\alpha=\pi$.
This singularity can be removed by subtracting the homogeneous solution with equivalent behavior.

For \textbf{even} $k$, the solution $\sum_l a_{kl} Y_{kl}$ of the homogeneous equation associated with the IFRR (\ref{114}) must be included into the general solution of the IFRR.
The coefficients $a_{kl}$ for bound states are determined by ensuring the wave function is normalisable at $r\rightarrow\infty$ (see, e.g., \cite{GM3}). Hence, these coefficients cannot be determined by analysis of the behavior of the wave function (\ref{1}) near the triple coalescence point.
The exception is the case of $k=2p$, when $h_{2p,p}\equiv0$ (see, e.g. \cite{AM1}).
Moreover, it was found \cite{MYE} that $a_{20}$ is identically zero (at least) for $^1S$ state by the required exchange symmetry of the spatial part of the wave function.
Otherwise, any admixture of $Y_{kl}$ preserves the correctness of $\psi_{k,p}^{(j)}$ as the finite solution of the IFRR (\ref{114}), making such a solution as the multi-valued one.
We present here a single-valued solutions, which can be produced by orthogonalization of the obtained
component $\psi_{k,p}^{(j)}$ to each of the $Y_{kl}$.
The resulting solutions can be called "pure" ones, because their HH expansions do not contain $Y_{kl}$
for any possible $l$.

\section{Previous Results}\label{S2}

In \cite{AM1}, the angular Fock coefficients were derived for the general Coulomb potential.
For the case of the helium isoelectronic sequence the following AFC become:
\begin{equation}\label{203}
\psi_{1,0}(\alpha,\theta))=-Z\varsigma+\frac{1}{2}\xi,
\end{equation}
\begin{equation}\label{204}
\psi_{2,1}(\alpha,\theta)=-Z\left(\frac{\pi-2}{3\pi}\right)\sin \alpha \cos \theta,
\end{equation}
\begin{equation}\label{205}
\psi_{3,1}(\alpha,\theta)=\frac{Z(\pi-2)}{36 \pi}
\left[6Z\varsigma\sin \alpha \cos \theta-
\xi\left(6-5\xi^2 \right)
\right],
\end{equation}
\begin{equation}\label{206}
\psi_{4,2}(\alpha,\theta)=\frac{(\pi-2)(5\pi-14)}{180 \pi^2}Z^2 \left(1-2 \sin^2\alpha \sin^2\theta\right),
\end{equation}
where
\begin{equation}\label{207}
\xi_1\equiv\frac{r_1}{r}=\cos\left(\frac{\alpha}{2}\right),~~~\xi_2\equiv\frac{r_2}{r}=\sin\left(\frac{\alpha}{2}\right),~~~
\varsigma=\xi_1+\xi_2,
\end{equation}
\begin{equation}\label{208}
\xi\equiv\frac{r_{12}}{r}=\sqrt{1-\sin \alpha \cos \theta},
\end{equation}
Using \emph{Mathematiaca}, we have verified that the AFC presented above satisfy the FRR (\ref{103}).

The derivation of the AFC, $\psi_{k,0}$ presents the most complicated problem. For $k=2$ this problem was successively solved, first of all, in the works \cite{AM1,GAM2,GM3}, where the $S$ states of different symmetry were presented in the natural $\left\{r_1,r_2,r_{12}\right\}$ coordinates for the general Coulomb potential. The expression for $\psi_{2,0}$ in the, so called, Pluvinage coordinates $\left \{\zeta,\eta \right \}$ was presented (without derivation) in \cite{FOR}. In the work \cite{LEZ1} the $\left \{\zeta,\eta \right \}$-representation from \cite{FOR} was transformed into the $\left\{r_1,r_2,r_{12}\right\}$ coordinates.
We have checked  both expressions for $\psi_{2,0}$, and found that they are, unfortunately, not correct, because they do not satisfy neither FRR (\ref{103}) nor the recursion relation expressed in the $\left \{\zeta,\eta \right \}$ coordinates (see, Eqs.(29), (25) \cite{FOR}), whereas the results of \cite{GM3} do satisfy.
Therefore, we present here the closed form of the AFC, $\psi_{2,0}$ obtained from the results of \cite{GM3},
and adapted to the helium isoelectronic sequence.
The condensed form expressed in the hyperspherical angles $\{\alpha,\theta\}$ reads:
\begin{eqnarray}\label{238}
 \psi_{2,0}=\frac{1}{12}(1-2E)+\frac{Z}{6\pi}\left\{-2\pi y \cos \theta\ln(\varsigma+\xi)+\pi x \ln\left[\frac{(x+\varsigma\xi)^2}{\varsigma^2(\gamma+x)}\right]+\gamma(2\beta+\pi)+
 \right.
\nonumber~~~\\
\left.
\pi\left(y-4\varsigma\xi\right)+ x\beta\left[\ln\left(\frac{\gamma-x}{\gamma+x}\right)+\ic(2\alpha-\pi)\right]+
x \alpha \ln\left(\frac{1+\cos \theta}{1-\cos \theta}\right)+\ic x \mathcal{L}
\right\}+
\nonumber~~~\\
Z^2\left(\frac{1}{2}y+\frac{1}{3}\right),~~~~~~
\end{eqnarray}
where
\begin{equation}\label{241}
 x=\cos \alpha,~~~~~~~y=\sin \alpha,
\end{equation}
\begin{equation}\label{239}
 \beta=\arcsin(\sin\alpha\cos\theta),~~~~~~\gamma=\xi\sqrt{2-\xi^2},
\end{equation}
\begin{equation}\label{240}
 \mathcal{L}=Li_2\left[e^{\ic(\alpha-\beta)}\right]+Li_2\left[-e^{-\ic(\alpha-\beta)}\right]-
 Li_2\left[-e^{-\ic(\alpha+\beta)}\right]-Li_2\left[e^{\ic(\alpha+\beta)}\right].
\end{equation}
Here $Li_2$ is the dilogarithm function, and $\ic=\sqrt{-1}$. We used the most convenient representation
\begin{equation*}
 L(\phi)=\frac{\ic}{2}\left[Li_2\left(e^{2\ic \phi}\right)-Li_2(1)-\phi(\phi-\pi)\right]
\end{equation*}
for the Lobachevsky function $L(\phi)$, which is valid for $0\leq\phi\leq\pi$ \cite{LL1,LL2}.

In the next sections we shall solve the IFFR (\ref{114}) for the components $\psi_{k,p}^{(j)}$ of the AFC defined by Eq.(\ref{112}).
We propose special methods for solving Eq.(\ref{114}) with different kinds of its rhs defined by Eq.(\ref{113}) and Eq.(\ref{103b}).
First, we show that these methods allow us to derive the correct expressions for the components of the AFC (\ref{203})-(\ref{206}) obtained previously.
Then, we fix the uncorrect representation for $\psi_{2,0}$ obtained in \cite{AM1}.
And at last, we derive the components of the AFC $\psi_{4,1}$ and $\psi_{3,0}$ that were not obtained previously.

\section{Technique for solving the IFRR with simplest nonseparable rhs}\label{S3}

In this section we discuss the solution of Eq.(\ref{114}) with the rhs $h_{k,p}^{(j)}$ presented by some polynomial in variable
$\xi\equiv\xi(\alpha,\theta)$ defined by Eq.(\ref{208}).
It follows from Eq.(\ref{103b}), that $h_{1,0}=-2V$, and hence $h_{1,0}^{(0)}=-2/\xi$
presents the simplest example of the rhs mentioned above.
We shall see that $h_{3,1}^{(1)}, h_{3,0}^{(0)}, h_{4,1}^{(1)}$, and many others are the examples of the rhs of that kind.
It is clear that a physical solution of the corresponding equation (\ref{114}) reduces to a function $\Phi(\xi)$ of a single variable $\xi$.
For example, it follows from Eq.(\ref{203}) that $\psi_{1,0}^{(0)}=\xi/2$.
It can be shown that the result of the direct action of the HAM (\ref{104}) to continues function $\Phi(\xi)$ is
\begin{equation}\label{310}
\Lambda^2\Phi(\xi)=\left(\xi^2-2\right)\Phi''(\xi)+\frac{5\xi^2-4}{\xi}\Phi'(\xi).
\end{equation}
Then, Eq.(\ref{114}) for $\psi_{k,p}^{(j)}(\alpha,\theta)\equiv\Phi_k (\xi)$ can be rewritten in the form
\begin{equation}\label{311}
\left(\xi^2-2\right)\Phi_k''(\xi)+\frac{5\xi^2-4}{\xi}\Phi_k'(\xi)-k(k+4)\Phi_k(\xi)=\textsl{h}(\xi),
\end{equation}
where  $h_{k,p}^{(j)}(\alpha,\theta)\equiv\textsl{h}(\xi)$.
The general solution of the homogeneous equation associated with Eq.(\ref{311}) can be presented in the form
\begin{equation}\label{312}
\Phi_k^{(h)}(\xi)=c_1 \textsl{u}_k(\xi)+c_2\textsl{v}_k(\xi),
\end{equation}
where the linearly independent solutions
\begin{equation}\label{313}
 \textsl{u}_k(\xi)=\frac{P_{k+3/2}^{1/2}\left(\xi/\sqrt{2}\right)}{\xi\sqrt[4]{2-\xi^2}},~~~~~~~~~~~~
\textsl{v}_k(\xi) =\frac{Q_{k+3/2}^{1/2}\left(\xi/\sqrt{2}\right)}{\xi\sqrt[4]{2-\xi^2}}
\end{equation}
are expressed via the associated Legendre functions, $P_\nu^\mu(x)$ and $Q_\nu^\mu(x)$ of the first and second kind, respectively.
The particular solution of the inhomogeneous equation (\ref{311}) can be found by the method of variation of parameters, that yields
\begin{equation}\label{314}
\Phi_k^{(p)}(\xi)=\textsl{v}_k(\xi)\int^\xi \frac{\textsl{u}_k(\xi')\textsl{h}(\xi')}{\left(\xi'^2-2\right)\mathcal{W}_k(\xi')}d\xi'-
\textsl{u}_k(\xi)\int^\xi \frac{\textsl{v}_k(\xi')\textsl{h}(\xi')}{\left(\xi'^2-2\right)\mathcal{W}_k(\xi')}d\xi',
\end{equation}
where the Wronskian has a simple form
\begin{equation}\label{315}
\mathcal{W}_k(\xi)=\frac{\sqrt{2}(k+2)}{\xi^2 \left(2-\xi^2\right)^{3/2}}.
\end{equation}
It was mentioned in Sec. \ref{S1} that the general solution of Eq.(\ref{311}) presents a sum of solutions (\ref{312}) and (\ref{314}).
The coefficients $c_1$ and $c_2$ in Eq.(\ref{312}) must be chosen in such a way that the final solution becomes the physically acceptable one.
It may be verified that $\textsl{u}_k(\xi)$ is divergent, whereas $\textsl{v}_k(\xi)$ is finite at the point $\xi=\sqrt{2}~(\alpha=\pi/2,\theta=\pi)$ for all integer $k$.
On the other hand, at the point $\xi=0~(\alpha=\pi/2,\theta=0)$
$\textsl{u}_k(\xi)$ is divergent for \textbf{even} values of $k$, whereas
$\textsl{v}_k(\xi)$ is divergent for \textbf{odd}  $k$.
This implies the following conclusions. For odd values of $k$, one should put $c_1=c_2=0$ if the particular solution (\ref{314}) satisfies the finiteness condition \textbf{P2}, otherwise the coefficients $c_1$ and$/$or $c_2$ must be chosen in such a way that to remove the divergence.
For even values of $k$, the additional condition of orthogonality of the final solution to $Y_{kl}(\alpha,\theta)$ enable us to obtain the "pure" solutions (see, the end of Sec. \ref{S1}).
In some complicated cases (see, e.g., Appendix \ref{SC}) the coupling equation (\ref{546}) can be applied.

Eq.(\ref{114}) with all of the right-hand sides mentioned at the beginning of this section, can be solved by the method described above.
However, for the most of the rhs, one can apply the more simple technique described below.

Substituting $\Phi(\xi)=B \xi^n$ into Eq.(\ref{310}), one obtains the following relation
\begin{equation*}
\Lambda^2 B \xi^n=B n \xi^n\left[n+4-2(n+1)\xi^{-2}\right],
\end{equation*}
where $B$ is arbitrary constant. Whence, one obtains
\begin{equation}\label{322}
\left[\Lambda^2-k(k+4)\right] B \xi^n=B\xi^n\left[(n-k)(n+k+4)-2n(n+1)\xi^{-2}\right].
\end{equation}
Given Eq.(\ref{322}), the particular solution (satisfying the finiteness condition) of the corresponding equation
\begin{equation}\label{323}
\left[\Lambda^2-k(k+4)\right]\Phi_k(\xi)=\textsl{h}(\xi)
\end{equation}
can be found in the form
\begin{equation}\label{324}
\Phi_k(\xi)=\sum_{i=0}^{i_h-1}B_i \xi^{2i+i_0},
\end{equation}
where $i_0=1$ for odd $n$, $i_0=0$ for even $n$,
and $i_h$ equals the number of terms in the polynomial presenting $\textsl{h}(\xi)$.
The unknown coefficients, $B_ i$ can be determined by means of substituting (\ref{324}) into Eq.(\ref{323}), using of Eq.(\ref{322}), and subsequent equating the coefficients for the same powers of $\xi$.

The FRR are solved in order of increasing $k$ and decreasing $p$. We shall calculate the AFC following this rule.
Thus, putting $\psi_{0,0}=1$, the FRR (\ref{103a}) for $k=1$, and $p=0$ is separated into two ones
\begin{equation}\label{325}
\left( \Lambda^2-5\right)\psi_{1,0}^{(j)}=h_{1,0}^{(j)},~~~~~~~~~~(j=0,1)
\end{equation}
where $h_{1,0}^{(0)}=-2/\xi$ (see, the beginning of this section).
The use of Eqs.(\ref{322})-(\ref{324}), enables us to calculate the component of the AFC,
\begin{equation}\label{326}
\psi_{1,0}^{(0)}=\frac{1}{2}\xi.
\end{equation}
This result is certainly consistent with Eq.(\ref{203}).
Note that all solutions of Eq.(\ref{114}) corresponding to the rhs, $\textsl{h}(\xi)$ (excepting $\psi_{3,0}^{(1c)}$ which is treated in Sec.\ref{S6a}) can be calculated by the simplified method presented by Eq.(\ref{322})-(\ref{324}).

\section{Technique for solving the IFRR with separable rhs}\label{S4}

This section is devoted to the method of solving Eq.(\ref{114}) with the rhs presented by the product of functions,
each of them depending on only one of the angle variables.

According to Eq.(\ref{103b}), the rhs of Eq.(\ref{325}) with $j=1$  has a form:
\begin{equation}\label{327}
h_{1,0}^{(1)}=2\left[\csc(\alpha/2)+\sec(\alpha/2)\right].
\end{equation}
A lot of the components of the rhs, presented by Eq.(\ref{103b}), have a form of the product
\begin{equation}\label{328}
h_{k,p}^{(j)}(\alpha,\theta)=P_l(\cos \theta)(\sin \alpha)^l \textmd{h}(\alpha),
\end{equation}
that for $l=0$ reduces to the function of a single variable $\alpha$, similar to Eq.(\ref{327}).
For convenience, we have introduced the notation $\textmd{h}(\alpha)\equiv h_{k,p}^{(j)}(\alpha)$.
To derive the corresponding component of the AFC, we propose the technique described below.

It can be shown \cite{GM3} that
\begin{equation}\label{329}
\Lambda^2 P_l(\cos \theta)f(\alpha)=-4 P_l(\cos \theta)\left[\frac{\partial^2}{\partial\alpha^2}
+2\cot \alpha  \frac{\partial}{\partial\alpha}-\frac{l(l+1)}{\sin^2\alpha}\right]f(\alpha).
\end{equation}
Hence, the solution of Eq.(\ref{114}) with $\psi_{k,p}^{(j)}=P_l(\cos \theta)f(\alpha)$ reduces to finding the function $f(\alpha)$ as a suitable solution of equation
\begin{equation}\label{330}
\left[4 \frac{\partial^2}{\partial\alpha^2}
+8\cot \alpha  \frac{\partial}{\partial\alpha}-\frac{4l(l+1)}{\sin^2\alpha}+k(k+4)\right]f(\alpha)=-(\sin \alpha)^l \textmd{h}(\alpha).
\end{equation}
Putting $f(\alpha)=(\sin \alpha)^l \textmd{g}(\alpha)$, one obtains the following equation
\begin{equation}\label{331}
4\textmd{g}''(\alpha)+8(l+1)\cot \alpha~ \textmd{g}'(\alpha)+(k-2l)(k+2l+4)\textmd{g}(\alpha)=-\textmd{h}(\alpha)
\end{equation}
for the function $\textmd{g}(\alpha)\equiv g_{k,p}^{(j)}(\alpha)$. The required solution of Eq.(\ref{114}) then becomes:
\begin{equation}\label{332}
\psi_{k,p}^{(j)}(\alpha,\theta)=P_l(\cos \theta)(\sin \alpha)^l\textmd{g}(\alpha).
\end{equation}
To solve Eq.(\ref{331}) it is convenient to make the change of variable,
\begin{equation}\label{334}
\rho=\tan(\alpha/2),
\end{equation}
which coincides with definition $\rho=(1-|x|)/y$ (given by (A11) \cite{AM1}) for $0\leq\alpha\leq\pi/2$, where $x$ and $y$ are defined by Eq.(\ref{241}).
Turning to the variable $\rho$, one obtains the following differential equation for the function  $g(\rho)\equiv\textmd{g}(\alpha)$, instead of Eq.(\ref{331}),
\begin{equation}\label{335}
\left(1+\rho^2\right)^2g''(\rho)+2\rho^{-1}\left[1+\rho^2+l(1-\rho^4)\right]g'(\rho)+(k-2l)(k+2l+4)g(\rho)=-h(\rho),
\end{equation}
where $h(\rho)\equiv \textmd{h}(\alpha)$.
Using the method of variation of parameters, one obtains the particular solution of Eq.(\ref{335}) in the form
\begin{equation}\label{340}
g(\rho)=v_{kl}(\rho)\int_{\rho_c}^\rho \frac{u_{kl}(\rho')h(\rho')}{(1+\rho'^2)^2 W_l(\rho')}d\rho' -
u_{kl}(\rho)\int_0^\rho \frac{v_{kl}(\rho')h(\rho')}{(1+\rho'^2)^2 W_l(\rho')}d\rho',
\end{equation}
where the linearly independent solutions of the homogeneous equation associated with Eq.(\ref{335}) are:
\begin{subequations}\label{342}
\begin{align}
u_{kl}(\rho)=\rho^{-2l-1}(\rho^2+1)^{\frac{k}{2}+l+2}~_2F_1\left(\frac{k+3}{2},\frac{k}{2}-l+1;\frac{1}{2}-l;-\rho^2\right),~~~~~~~\label{342a}\\
v_{kl}(\rho)=(\rho^2+1)^{\frac{k}{2}+l+2}~_2F_1\left(\frac{k+3}{2},\frac{k}{2}+l+2;l+\frac{3}{2};-\rho^2\right).~~~~~~~~~~~~~~~~\label{342b}
\end{align}
\end{subequations}
Here $_2F_1$ is the Gauss hypergeometric function.
It is important to note that:

1) The corresponding Wronskian
\begin{equation}\label{343}
W_l(\rho)=-\frac{2l+1}{\rho}\left(\frac{\rho^2+1}{\rho}\right)^{2l+1}
\end{equation}
is independent of $k$;

2) For the case of $k=2l$, the solution $v_{2l,l}(\rho)=1$ (follows directly from Eq.(\ref{335})).

The lower limits of integration in Eq.(\ref{340}) must be chosen in such a way as to remove singularities by subtracting the homogeneous solutions with equivalent behavior. This yields $\rho_c=1$ for even $k$, and $\rho_c=\infty$ for odd $k$.

One should emphasize that, in fact, functions (\ref{342}) are presented by a rather simple elementary functions.
For the particular case of the rhs (\ref{327}) corresponding to $k=1,p=0,l=0,j=1,\rho_c=\infty$, one obtains:
\begin{equation}\label{344}
h(\rho)=\frac{2(1+\rho)\sqrt{1+\rho^2}}{\rho},~~~~~u_{10}(\rho)=\frac{1-3\rho^2}{\rho\sqrt{1+\rho^2}},
~~~~~v_{10}(\rho)=\frac{3-\rho^2}{3\sqrt{1+\rho^2}}.~~~~~~~~~~
\end{equation}
Application of the form (\ref{340}) yields for this case:
\begin{equation}\label{345}
\psi_{1,0}^{(1)}=-\frac{1+\rho}{\sqrt{1+\rho^2}}=-\left[\sin(\alpha/2)+\cos(\alpha/2)\right],
\end{equation}
what certainly corresponds to Eq.(\ref{203}).

\section{Explicite solution for  $\psi_{2,p}$}\label{S5}

For the case of $k=2,~p=1$, the AFC is presented by Eq.(\ref{204}).
Derivation of the AFC, $\psi_{k,p}$ with $k=2p$ is a very simple task. Its simplest solution was detailed in Sec.4.3 of the paper \cite{AM1}.

According to (\ref{103}), the FRR for $k=2,~p=0$ is
\begin{equation}\label{401}
(\Lambda^2-12)\psi_{2,0}=8\psi_{2,1}-2 V \psi_{1,0}+2E.
\end{equation}
Using Eqs.(\ref{105}), (\ref{203}) and (\ref{204}), one can present Eq.(\ref{401}) in components
\begin{equation}\label{402}
(\Lambda^2-12)\psi_{2,0}^{(j)}=h_{2,0}^{(j)},~~~~~~~~~~~~~~(j=0,1,2)
\end{equation}
where
\begin{subequations}\label{403}
\begin{align}
h_{2,0}^{(0)}=2E-1,~~~~~~~~~~~~~~~~~~~~~~~~~~~~~~~~~~~~~~~~~~~~~~~~~~~~~~~\label{403a}\\
h_{2,0}^{(1)}=h_{2,0}^{(1a)}+h_{2,0}^{(1b)},~~~~~~~~~~~~~~~~~~~~~~~~~~~~~~~~~~~~~~~~~~~~~~~~~\label{403b}\\
h_{2,0}^{(2)}=-4(1+\csc \alpha),~~~~~~~~~~~~~~~~~~~~~~~~~~~~~~~~~~~~~~~~~~~~~~\label{403c}
\end{align}
\end{subequations}
\begin{subequations}\label{404}
\begin{align}
h_{2,0}^{(1a)}=\frac{2\left[\sin(\alpha/2)+\cos(\alpha/2)\right]\left(2\csc \alpha-3\cos \theta+3\right)}
{3\sqrt{1-\sin \alpha \cos \theta}},~~~~~\label{404a}\\
h_{2,0}^{(1b)}=\frac{\csc(\alpha/2)+\sec(\alpha/2)}{3\sqrt{1-\sin \alpha \cos \theta}}-\frac{8(\pi-2)\sin \alpha \cos \theta}{3\pi}.~~~~~~~~~\label{404b}
\end{align}
\end{subequations}
Using the technique presented in Sec.\ref{S4}, one obtains :
\begin{equation}\label{405}
u_{20}(\rho)=\frac{1}{\rho}+\rho-\frac{8\rho}{1+\rho^2},~~~~~~~~~~~~v_{20}(\rho)=\frac{1-\rho^2}{1+\rho^2}.
\end{equation}
Application of the form (\ref{340}) for $k=2,p=0,l=0,j=0,\rho_c=1$ yields
\begin{equation}\label{406}
h(\rho)=2E-1,~~~~~~~~~~\psi_{2,0}^{(0)}=\frac{1}{12}(1-2E).
\end{equation}
Similarly, for $k=2,p=0,l=0,j=2,\rho_c=1$, one obtains:
\begin{equation}\label{407}
h(\rho)=-\frac{2(1+\rho)^2}{\rho},~~~~~~~~~~\psi_{2,0}^{(2)}=\frac{1}{3}+\frac{\rho}{1+\rho^2}=\frac{1}{3}+\frac{1}{2}\sin \alpha.
\end{equation}

\subsection{Specific solution for $\psi_{2,0}^{(1)}$ }\label{S5b}

In Sec. \ref{S2}, we presented (without derivation) the closed form of $\psi_{2,0}$ obtained from the results of the work \cite{GM3}.
This form defined by Eq.(\ref{238}) is very convenient for presentation $\psi_{2,0}$ itself.
However, it includes functions (e.g., dilogarithm function with argument in the form of the exponential function) which are too complex for further mathematical processing required for derivation of the higher-order angular Fock coefficients.
For example, according to Eq.(\ref{103b}) the rhs, $h_{3,0}$ of the FRR (\ref{103a}) for the AFC $\psi_{3,0}$, contains the term $-2V \psi_{2,0}$.
To obtain the particular solution of the corresponding FRR, one needs to apply the integral formula (\ref{340}) with integrands containing $h_{3,0}$.
It is clear that the reduction of the required integration to the analytic form is impossible,
whereas the representation of $\psi_{2,0}$ in the form of infinite single/double series enables us to provide this integration.
Such series were presented in \cite{AM1}. The problem is that, unfortunately, there are too many errors/misprints in the important final formulas, whereas the methods of its obtaining have been described rather superficially or not at all been described.

In this paper, we rederive these formulas, while describing in detail our method that differs from that of \cite{AM1}.

It is shown in Appendix \ref{SA} that the following representation is valid:
\begin{equation}\label{541}
\psi_{2,0}^{(1)}=-\frac{1}{3}\left[\sin(\alpha/2)+\cos(\alpha/2)\right]\sqrt{1-\sin \alpha \cos\theta}+\chi_{20}(\alpha,\theta),
\end{equation}
where the HH expansion for the function $\chi_{20}$ reads
\begin{equation}\label{542}
\chi_{20}(\alpha,\theta)=\frac{2}{3}{\sum_{nl}}'\frac{D_{nl}}{(n-2)(n+6)}Y_{nl}(\alpha,\theta),
\end{equation}
and $D_{nl}$ is defined by Eq.(\ref{A21}).
The prime denotes that $n=2$ must be omitted from summation.
Note that (\ref{542}) requires a double summation (over $n$ and $l$), which possesses a slow convergence.\\
In \cite{AM1} the following single series representation was proposed :
\begin{equation}\label{543}
\chi_{20}(\alpha,\theta)=\sum_{l=0}^\infty P_l(\cos\theta)(\sin \alpha)^l\sigma_l(\alpha).
\end{equation}
The problem is that the technique, used for deriving the functions $\sigma_l$, is complex and ambiguous, whereas the final formulas ((\emph{A}19), (\emph{A}22) and (\emph{A}24) from \cite{AM1}) were presented with errors/misprints.
We present here the alternative method to derive $\sigma_l$.

Suppose that a regular function having the unnormalized HH expansion (\ref{110}), can be presented in the form of infinite single series:
\begin{equation}\label{545}
\mathcal{F}(\alpha,\theta)=\sum_{l=0}^\infty P_l(\cos \theta)Q_l(\alpha).
\end{equation}
Multiplication of the right-hand sides of Eq.(\ref{110}) and Eq.(\ref{545}) by $Y_{2l',l'}\equiv(\sin\alpha)^{l'} P_{l'}(\cos \theta) $, and subsequent integration over the angular space (\ref{109}), enables one to obtain the following coupling equation:
\begin{equation}\label{546}
\mathcal{F}_{2l,l}=\frac{(l+1)!}{\sqrt{\pi}\Gamma(l+3/2)}\int_0^\pi Q_l(\alpha)(\sin \alpha)^{l+2}d\alpha.
\end{equation}
We used Eq.(\ref{107}), the orthogonality equation (\ref{108}) for HH's, and the well-known formula of orthogonality for the Legendre polynomials.
Putting $Q_l(\alpha)=(\sin\alpha)^l \sigma_l(\alpha)$, where $\sigma_l(\alpha)$ is included into Eq.(\ref{543}),
then using expansion (\ref{542}) and representation (\ref{A21}), and simplifying, one obtains instead of the latter relation:
\begin{eqnarray}\label{547}
3(l-1)(l+1)(l+3)\int_0^\pi (\sin \alpha)^{2l+2}\sigma_l(\alpha)d\alpha-1=
\nonumber~~~~~~~~~~~~~~~~~~~~~~~~~~~~~~~~~~~~~~~~~~~~~~~~~~~~~~~\\
\frac{\sqrt{\pi} \Gamma(l+3/2)}{2^{l+2}l!}~_3F_2\left(\frac{l+1}{2},\frac{l}{2}+1,l+\frac{3}{2};l+2,l+2;1\right).~~~~~~~~~
\end{eqnarray}
Note that the case of $l=1$ cannot be used in Eq.(\ref{547}), because the term with $Y_{21}$  is excluded from the HH expansion  (\ref{542}).
It is clear that for $l=1$ the integral in the lhs of Eq.(\ref{547}) just equals zero.

Application of the relations (\ref{330})-(\ref{335}) and (\ref{547}) enables us (details can be found in Appendix \ref{SB}) to derive the following representations which are valid in the range $\alpha\in[0,\pi/2]$:
\begin{equation}\label{548}
\sigma_0=\frac{1}{12}\left\{\left(2y-\frac{1}{y}\right)\alpha+|x|
\left[1+ 2\ln(|x|+1) \right]-y-2 \right\},~~~~~~~~~~~~~~~~~~~~~~~~~~~~~~~~~~~~~~~~~~~~
\end{equation}
\begin{eqnarray}\label{549}
\sigma_1=\frac{1}{24\pi}\left[\left(\frac{1}{\rho^3}+\frac{9}{\rho}-9\rho-\rho^3\right)\arctan \rho -
\frac{1}{\rho^2}-\rho^2-\pi-\frac{8}{3}+16 G\right]-
\nonumber~~~~~~~~~~~~~~~~~~~~~~\\
\frac{\rho\left(\rho^2-6\rho+3\right)}{72}+
\frac{1}{6}\ln\left(\frac{1+\rho^2}{4}\right).~~~~~~~~~~~
\end{eqnarray}
\begin{eqnarray}\label{550}
\sigma_l=-\frac{2^{-l-1}}{3}\left\{(1+\rho^2)^{l-1}\left[\frac{l\rho^3}{(l+1)(l+2)}-\frac{\rho^2}{l}+\frac{\rho}{l+1}-\frac{l+1}{l(l-1)}\right]+
\right.
\nonumber~~~~~~~~~~~~~~~~~~~~~~~~~~\\
\left.
\frac{(l-2)!\Gamma\left(\frac{l+1}{2}\right)}{\Gamma\left(l+\frac{1}{2}\right)\Gamma\left(\frac{l}{2}+1\right)}
~_2F_1\left(\frac{l-1}{2},\frac{l+3}{2};l+\frac{3}{2};y^2\right)\right\},~~~~~~~~~~~~~~~~~~~~~~~(l\geq2)~~~~~
\end{eqnarray}
where $G\simeq 9159656$ is the Catalan's constant, and $x,y$ and $\rho$ are defined by Eq.(\ref{241}) and Eq.(\ref{344}), respectively.
For $\pi/2<\alpha\leq \pi$, one needs to replace $\alpha$ by $\pi-\alpha$, $x$ by $-x$, and $\rho$ by $1/\rho$.
One can optionally to set $\rho=(1-|x|)/y$, which is valid for the whole range $\alpha\in[0,\pi]$.
Note that factor $1/4$ is missed in the corresponding expression (\emph{A}24) from \cite{AM1}.
Moreover, one should emphasize that representation (\emph{A}22) from \cite{AM1} for $\sigma_1$ is not correct, because it doesn't satisfy the inhomogeneous differential equation (\ref{B8}), and doesn't agree with definition (\ref{551}) presented below.

Using series rearrangement (see, \cite{AM1}, \cite{RED}) of the double summation in Eq.(\ref{542}), one can obtain another representation
\begin{equation}\label{551}
\sigma_l=\frac{1}{6}\sum_{\delta_{l,1}}\frac{D_{4m+2l,l}}{(2m+l-1)(2m+l+3)}C_{2m}^{(l+1)}(x),
\end{equation}
which is valid for any $l\geq0$.  The convergence of expansion (\ref{551}) is very slow, however, it can be used to verify the correctness of Eqs.(\ref{548})-(\ref{550}).

It is seen that the component  $\psi_{2,0}^{(1)}$ is presented by Eqs.(\ref{541})-(\ref{542}).
On the other hand, using definition (\ref{112}), one can separate the component  $\widetilde{\psi}_{2,0}^{(1)}$ out of the AFC $\psi_{2,0}$ defined by Eq.(\ref{238}). We have intentionally put up a sign "tilde" over $\psi$, because the components obtained by these two different methods
are not coincident for any point of the angular space $\left\{\alpha,\theta\right\}$.
It is because both representations mentioned above are not the "pure" solutions (see, the end of Sec.\ref{S1}) of Eq.(\ref{402}).
In other words, the HH expansions of both components include some admixture of the homogeneous solutions $Y_{2,l}$ ($l=0,1$) of Eq.(\ref{402}).
It is clear that $\chi_{20}$ cannot contain such an admixture by definition (\ref{542}).
Thus, using definitions (\ref{541}) and (\ref{111}), one obtains the coefficient
\begin{equation*}
C_{21}^{(p)}=N_{21}^2\int Y_{21}(\alpha,\theta)\left\{-\frac{1}{3}\left[\sin(\alpha/2)+\cos(\alpha/2)\right]\sqrt{1-\sin \alpha \cos \theta}\right\}d\Omega=\frac{\pi+4}{9\pi},
\end{equation*}
for the unnormalized HH, $Y_{21}(\alpha,\theta)=\sin \alpha \cos \theta$ in the HH expansion of  $\psi_{2,0}^{(1)}$.
It can be verified that $C_{20}^{(p)}=0$. Thus, one obtains  the "pure" component in the form
$\psi_{2,0}^{(1p)}=\psi_{2,0}^{(1)}-C_{21}^{(p)}\sin \alpha \cos \theta$.
It is clear that another way to obtain the "pure" component $\psi_{2,0}^{(1p)}$ is to subtract $C_{21}^{(p)}$ from $\sigma_1(\alpha)$ defined by Eq.(\ref{549}).
The same method can be used to obtain the "pure" component $\widetilde{\psi}_{2,0}^{(1p)}=\widetilde{\psi}_{2,0}^{(1)}-\widetilde{C}_{21}^{(p)}\sin \alpha \cos \theta$ based on the analytic expression (\ref{238}). Numerical integration yields the following value of the coefficient
\begin{equation*}
\widetilde{C}_{21}^{(p)}=N_{21}^2\int Y_{21}(\alpha,\theta)\widetilde{\psi}_{2,0}^{(1p)}(\alpha,\theta)d\Omega=0.315837352,
\end{equation*}
whereas $\widetilde{C}_{20}^{(p)}=0$.
The "pure" components $\psi_{2,0}^{(1p)}$ and $\widetilde{\psi}_{2,0}^{(1p)}$ are certainly coincident for any angular point under consideration.

\section{Solutions for AFC with $k>2$}\label{S6}

For $k=3$ and $p=1$, the FRR (\ref{103}) reduces to
\begin{equation}\label{601}
\left[\Lambda^2-21\right]\psi_{3,1}=-2V\psi_{2,1},
\end{equation}
where $V$ and $\psi_{2,1}$ are defined by Eq.(\ref{105}) and Eq.(\ref{204}), respectively.
According to equations (\ref{112})-(\ref{114}), the considered Eq.(\ref{601}) can be presented in components
\begin{equation}\label{602}
\left[\Lambda^2-21\right]\psi_{3,1}^{(j)}=h_{3,1}^{(j)},~~~~~~~~~~~(j=1,2)
\end{equation}
where
\begin{subequations}\label{603}
\begin{align}
h_{3,1}^{(1)}=2B(\frac{1}{\xi}-\xi),~~~~~~~~~~~~~~~~~~~~~~~~~~~~~~~~~~\label{603a}\\
h_{3,1}^{(2)}=-4B \cos \theta[\sin(\alpha/2)+\cos(\alpha/2)],~~~~~~~~~\label{603b}
\end{align}
\end{subequations}
with constant $B=(\pi-2)/3\pi$, and $\xi$ defined by Eq.(\ref{208}).
Using the technique described in Sec.\ref{S3}, one easily find the component
\begin{equation}\label{604}
\psi_{3,1}^{(1)}=B\xi\left(\frac{5}{12}\xi^2-\frac{1}{2}\right).
\end{equation}
To solve Eq.(\ref{602}) with $j=2$, one can apply the technique described in Sec.\ref{S4}. Thus, for the case of $k=3,p=1,l=1,\rho_c=\infty$, one obtains:
\begin{equation}\label{605}
h(\rho)=-2B\frac{(\rho+1)\sqrt{\rho^2+1}}{\rho},~~~u_{3,1}(\rho)=\frac{1+14\rho^2-35\rho^4}{\rho^3\sqrt{1+\rho^2}},
~~~v_{3,1}=\frac{35-14\rho^2-\rho^4}{35\sqrt{1+\rho^2}}.~~~
\end{equation}
Application of the formula (\ref{340}) then yields:
\begin{equation}\label{606}
g(\rho)=\frac{B(1+\rho)}{2\sqrt{1+\rho^2}},~~~~~\psi_{3,1}^{(2)}=g \sin \alpha\cos\theta=
\frac{B}{2}\left[\sin\left(\frac{\alpha}{2}\right)+\cos\left(\frac{\alpha}{2}\right)\right]\sin \alpha \cos \theta.
\end{equation}
The components presented by Eq.(\ref{604}) and Eq.(\ref{606}) are certainly consistent with AFC $\psi_{3,1}$ presented by Eq.(\ref{205}).

\subsection{The AFC previously undetermined} \label{S6a}

Let us consider the FRR for the following two cases: i) $k=3,p=0$, and ii) $k=4,p=1$. According to (\ref{103}) one obtains:
\begin{equation}\label{701}
\left(\Lambda^2-21\right)\psi_{3,0}=10\psi_{3,1}-2V\psi_{2,0}+2E\psi_{1,0},
\end{equation}
\begin{equation}\label{702}
\left(\Lambda^2-32\right)\psi_{4,1}=24\psi_{4,2}-2V\psi_{3,1}+2E\psi_{2,1}.
\end{equation}
Using Eqs.(\ref{112})-(\ref{114}) and the AFC previously determined, one can present the FRR (\ref{701}), (\ref{702}) in components, as follows:
\begin{equation}\label{703}
\left(\Lambda^2-21\right)\psi_{3,0}^{(j)}=h_{3,0}^{(j)},~~~~~~~~~~~~~~~~(j=0,1,2,3)
\end{equation}
\begin{equation}\label{704}
\left(\Lambda^2-32\right)\psi_{4,1}^{(j)}=h_{4,1}^{(j)},~~~~~~~~~~~~~~~~~(j=1,2,3)
\end{equation}
where
\begin{equation}\label{705}
h_{3,0}^{(0)}=2E\psi_{1,0}^{(0)}-2V_0\psi_{2,0}^{(0)}=E\xi+\frac{2E-1}{6\xi},~~~~~~~~~~~~~~~~~~~~~~~~~~~~~~~
\end{equation}
\begin{eqnarray}\label{706}
h_{3,0}^{(1)}=2E\psi_{1,0}^{(1)}+2V_1\psi_{2,0}^{(0)}-2V_0\psi_{2,0}^{(1)}+10\psi_{3,1}^{(1)}=
\nonumber~~~~~~~~~~~~~~~~~~~~~~~~~~~~\\
-\frac{2}{\xi}\chi_{20}(\alpha,\theta)+
\frac{5(\pi-2)(5\xi^3-6\xi)}{18\pi}+
\left(\frac{1-2E}{3}\right)\frac{\varsigma}{\sin \alpha}+
2\left(\frac{1}{3}-E\right)\varsigma
,~~~~~~~~~~~~~~~~~~~~
\end{eqnarray}
\begin{eqnarray}\label{707}
h_{3,0}^{(2)}=2V_1\psi_{2,0}^{(1)}-2V_0\psi_{2,0}^{(2)}+10\psi_{3,1}^{(2)}=
\nonumber~~~~~~~~~~~~~~~~~~~~~~~~~~~~~~~~~~~~~\\
\frac{4\varsigma}{\sin \alpha}\chi_{20}(\alpha,\theta)-
\frac{2}{3}\left(2\xi+\frac{1}{\xi}\right)-
\frac{4\xi}{3\sin \alpha}-\frac{\sin \alpha}{\xi}+
\frac{5(\pi-2)}{3\pi}\varsigma\sin \alpha \cos \theta,~~~~~
\end{eqnarray}
\begin{equation}\label{708}
h_{3,0}^{(3)}=2V_1\psi_{2,0}^{(2)}=
\frac{2}{3}\left(\frac{2}{\sin \alpha}+3\right)\varsigma,~~~~~~~~~~~~~~~~~~~~~~~~~~~~~~
\end{equation}
\begin{equation}\label{709}
h_{4,1}^{(1)}=2E\psi_{2,1}^{(1)}-2V_0\psi_{3,1}^{(1)}=\frac{\pi-2}{18\pi}
\left[6(1-2E)+(12E-5)\xi^2\right],~~~~~~~~~~~~~~~~~~~~~~~~~~~~~~~
\end{equation}
\begin{eqnarray}\label{710}
h_{4,1}^{(2)}=2V_1\psi_{3,1}^{(1)}-2V_0\psi_{3,1}^{(2)}+24\psi_{4,2}^{(2)}=\frac{\pi-2}{15\pi^2}\times
\nonumber~~~~~~~~~~~~~~~~~~~~~~~~~~~~~~~~~~~~~~~~~~~~~~~~~~\\
\left\{
2(5\pi-14)\left[1-\frac{4}{3}\sin^2\alpha+\frac{4}{3}\sin^2\alpha P_2(\cos \theta)\right]+
5\pi\varsigma\left[\frac{5}{3\sin \alpha}\xi^3+\left(1-\frac{2}{\sin\alpha}\right)\xi-\frac{1}{\xi}\right]
\right\},~~~
\end{eqnarray}
\begin{equation}\label{711}
h_{4,1}^{(3)}=2V_1\psi_{3,1}^{(2)}=\frac{\pi-2}{3\pi}\left[2+\tan\left(\frac{\alpha}{2}\right)+
\cot\left(\frac{\alpha}{2}\right)\right]\sin \alpha \cos \theta.
\end{equation}
Here, $V_0$ and $V_1$ are defined by relation $V=V_0-Z V_1$ and Eq.(\ref{105}), whereas $\varsigma,~\xi$  and $\chi_{20}$ are defined by Eqs.(\ref{207}), (\ref{208}) and Eq.(\ref{543}), respectively.

It is seen that the rhs $h_{3,0}^{(0)}$ and $h_{4,1}^{(1)}$ present functions of only $\xi$, and the corresponding components
$\psi_{3,0}^{(0)}$ and $\psi_{4,1}^{(1)}$ can be derived by simplified method described in Sec.\ref{S3} (see, Eqs.(\ref{322})-(\ref{324})).
The results are presented in Table \ref{T1}.
The components $\psi_{3,0}^{(3)}$ and $\psi_{4,1}^{(3)}$, obtained by technique described in Sec.\ref{S4},
are presented in Table \ref{T1}, as well.

Extending separation presented in Sec.\ref{S1}, one can obtain the solutions to the majority of subcomponents of the remaining components presented by
Eqs.(\ref{703}), (\ref{704}). In particular, let us perform the following additional separation,
\begin{equation}\label{751}
\psi_{3,0}^{(1)}=\psi_{3,0}^{(1a)}+\psi_{3,0}^{(1b)}+\psi_{3,0}^{(1c)}+\psi_{3,0}^{(1d)},
\end{equation}
\begin{equation}\label{752}
\psi_{3,0}^{(2)}=\psi_{3,0}^{(2a)}+\psi_{3,0}^{(2b)}+\psi_{3,0}^{(2c)}+\psi_{3,0}^{(2d)},
\end{equation}
\begin{equation}\label{753}
\psi_{4,1}^{(2)}=\psi_{4,1}^{(2b)}+\psi_{4,1}^{(2c)}+\psi_{4,1}^{(2d)}.
\end{equation}
Subcomponents $\psi_{3,0}^{(1a)}$ and $\psi_{3,0}^{(2a)}$ are calculated with using  Eqs.(\ref{322})-(\ref{324}), and are presented in Table \ref{T1} together with the corresponding rhs.
Subcomponents $\psi_{3,0}^{(1b)}$ and $\psi_{4,1}^{(2b)}$, obtained by the method described in Sec.\ref{S4},
are presented in Table \ref{T1} together with the corresponding rhs.
Note that expression for $\psi_{4,1}^{(2b)}$ presented in Table \ref{T1} is correct for $0\leq\alpha\leq\pi/2$, whereas for $\pi/2<\alpha\leq\pi$ one should replace $\alpha$ by $\pi-\alpha$.
The remaining rhs are:
\begin{equation}\label{754}
h_{3,0}^{(1c)}=\frac{25(\pi-2)}{18\pi}\xi^3,~~~~~~~h_{3,0}^{(1d)}=-\frac{2}{\xi}\chi_{20}(\alpha,\theta),
\end{equation}
\begin{equation}\label{755}
h_{3,0}^{(2b)}=\frac{5(\pi-2)}{3\pi}\varsigma \sin\alpha \cos \theta,~~~~~~
h_{3,0}^{(2c)}=-\frac{4\xi}{3\sin \alpha},~~~~~~
h_{3,0}^{(2d)}=\frac{4\varsigma}{\sin \alpha}\chi_{20}(\alpha,\theta),
\end{equation}
\begin{subequations}\label{756}
\begin{align}
h_{4,1}^{(2c)}=\frac{8(\pi-2)(5\pi-14)}{45\pi^2}\sin^2\alpha P_2(\cos \theta),~~~~~~~~~~~~~~~~~~~~~~~~~~~~~~~~~~~~~~~\label{756a}\\
h_{4,1}^{(2d)}=\frac{\pi-2}{3\pi}
\left[\sin\left(\frac{\alpha}{2}\right)+\cos\left(\frac{\alpha}{2}\right)  \right]\left[\frac{5}{3\sin \alpha}\xi^3+\left(1-\frac{2}{\sin\alpha}\right)\xi-\frac{1}{\xi}\right].~~~\label{756b}
\end{align}
\end{subequations}
To calculate the subcomponent $\psi_{3,0}^{(1c)}$, one can use the particular solution presented in Sec.\ref{S3}.
Putting $k=3,~\textsl{h}(\xi)=\xi^3$ in Eq.(\ref{314}), one obtains:
\begin{equation*}
\Phi_3^{(p)}(\xi)=\frac{1}{8}\left[\frac{\xi\left(3-5\xi^2\right)}{6}-
\frac{\left(4\xi^4-10\xi^2+5\right)\arcsin \left(\xi/\sqrt{2}\right)}{5\sqrt{2-\xi^2}}\right].
\end{equation*}
The problem is that $\Phi_3^{(p)}(\xi)$ is singular at the point $\xi=\sqrt{2}~(\alpha=\pi/2,\theta=\pi)$.
This singularity can be eliminated with the help of the function $\textsl{u}_3(\xi)$ (see, Eq.(\ref{313})) presenting a solution of the associated homogeneous equation, and having the same kind of singularity.
Thus, given that
\begin{equation*}
x_1=\lim_{\xi\rightarrow\sqrt{2}}\Phi_3^{(p)}(\xi)\sqrt{\xi-\sqrt{2}}=\frac{\ic \pi}{80\times2^{3/4}},~~
x_2=\lim_{\xi\rightarrow\sqrt{2}}\textsl{u}_3(\xi)\sqrt{\xi-\sqrt{2}}=-\frac{\ic}{\sqrt{2\pi}}~~~~~~
\end{equation*}
one obtains, finally
\begin{eqnarray}\label{757}
\psi_{3,0}^{(1c)}(\alpha,\theta)=\frac{25(\pi-2)}{18\pi}\left[\Phi_3^{(p)}(\xi)-\frac{x_1}{x_2}\textsl{u}_3(\xi)\right]=~~~~~~~~~~~~~~~~~~
\nonumber~~~~~~~~~~~~~~~~~~~~~~~~~~~~~~~~~~\\
\frac{25(\pi-2)}{144\pi}\left[\frac{\xi\left(3-5\xi^2\right)}{6}+
\frac{\left(4\xi^4-10\xi^2+5\right)\arccos \left(\xi/\sqrt{2}\right)}{5\sqrt{2-\xi^2}}\right].~~~~~~~~~~~~~~~~~
\end{eqnarray}
Using the technique presented in Sec.\ref{S4}, one obtains for subcomponents with the rhs $h_{3,0}^{(2b)}$ and $h_{4,1}^{(2c)}$ defined by Eq.(\ref{755}) and Eq.(\ref{756a}), respectively,
\begin{eqnarray}\label{714}
\psi_{3,0}^{(2b)}=\frac{(\pi-2)(1+\rho^2)^{-3/2}}{288\pi\rho^2}\times
\nonumber~~~~~~~~~~~~~~~~~~~~~~~~~~~~~~~~~~~~~~~~~~~~~~~~~~~~~~~~~~~~~~~~~~~~~\\
\left[\alpha-2\rho+14\alpha\rho^2-35\rho^3(\pi-\alpha+2)-35\rho^4(\alpha+2)+(14\rho^5+\rho^7)(\pi-\alpha)-2\rho^6\right]\cos\theta,~~~~~
\end{eqnarray}
\begin{eqnarray}\label{716}
\psi_{4,1}^{(2c)}=-\frac{(\pi-2)(5\pi-14)}{8640\pi^2}\sin^2\alpha P_2(\cos\theta)\times
\nonumber~~~~~~~~~~~~~~~~~~~~~~~~~~~~~~~~~~~~~~~~~~~~~~~~~~\\
\left\{\frac{187}{15}+\frac{1}{4\rho^5}
\left[\alpha\left(\rho^2-1\right)\left(3\rho^8+28\rho^6+178\rho^4+28\rho^2+3\right)+
6\rho\left(\rho^8+8\rho^6+8\rho^2+1\right)\right]
\right\},~~~
\end{eqnarray}
where $\rho$ is defined by Eq.(\ref{334}). Representations (\ref{714}) and (\ref{716}) are correct, as previously, for $\alpha\in[0,\pi/2]$. For $\pi/2<\alpha\leq\pi$ one should replace $\rho$ by $1/\rho$, and $\alpha$ by $\pi-\alpha$.
At first sight it may seem that the rhs of Eq.(\ref{714}) diverges as $\alpha\rightarrow 0$. However, it may be verified that it is wrong, and what is more $\psi_{3,0}^{(2b)}$ achieves zero as $\alpha\rightarrow 0$.

For subcomponent $\psi_{4,1}^{(2d)}$ corresponding to the rhs (\ref{756b})
we use the representation
\begin{equation}\label{718}
\psi_{4,1}^{(2d)}(\alpha,\theta)=\sum_{l=0}^\infty P_l(\cos\theta)(\sin \alpha)^l \textmd{t}_l(\alpha),
\end{equation}
similar to Eq.(\ref{543}). Function $\tau_l(\rho)\equiv \textmd{t}_l(\alpha)$ can be presented (details can be found in Appendix \ref{SC}) in the form:
\begin{eqnarray}\label{719}
\tau_{0}(\rho)=\frac{\pi-2}{108\pi\left(\rho^2+1\right)^2}
\left\{\frac{1}{15}\left(19\rho^5+75\rho^4-60\rho^3+30\rho^2+45\rho-45\right)+
\right.
\nonumber~~~~~~~~~~~~~~~~~~~~~~~~~~~~~\\
\left.
\left(\rho^5-15\rho^3+15\rho-\frac{1}{\rho}\right)\arctan \rho+
\left(3\rho^4-10\rho^2+3\right)
\left[\frac{247}{75\pi}-\frac{4G}{\pi}-\ln\left(\frac{\rho^2+1}{4}\right)\right]
\right\},~~~~~~~~~~
\end{eqnarray}
\begin{eqnarray}\label{720}
\tau_{1}(\rho)=-\frac{\pi-2}{302400\pi\rho^2\left(\rho^2+1\right)}
\left\{1268\rho^7-2505\rho^6+1960\rho^5+32263\rho^4+18900\rho^3+18305\rho^2-735+
\right.
\nonumber~\\
\left.
735\left[\left(\rho^7+20\rho^5-90\rho^3+20\rho+\frac{1}{\rho}\right)\arctan \rho-
32\rho^2(\rho^2-1)\ln\left(\frac{\rho^2+1}{2}\right)\right]
\right\},~~~~~~~~~
\end{eqnarray}
\begin{eqnarray}\label{721}
\tau_{2}(\rho)=\frac{\pi-2}{14175\pi\rho^4}
\left\{\frac{14}{\pi}\left(41-150 G+\frac{3765\pi}{128}\right)\rho^4+
\right.
\nonumber~~~~~~~~~~~~~~~~~~~~~~~~~~~~~~~~~~~~~~~~~~~~~~~~~~~~~~~~~~~\\
\frac{5}{128}\left(672\rho^9-465\rho^8+760\rho^7-6720\rho^6-5880\rho^5+2520\rho^2+315\right)+
\nonumber~~~~~~~~~~~~~~~~~~~~~~~~~~~~~\\
\left.
525\left[\frac{(\rho^2-1)}{128}\left(3\rho^7+28\rho^5+178\rho^3+28\rho+\frac{3}{\rho}\right)\arctan \rho-
\rho^4\ln\left(\frac{\rho^2+1}{4}\right)\right]
\right\},~~~~~~~~
\end{eqnarray}
\begin{eqnarray}\label{724}
\tau_{l}(\rho)=\frac{(\pi-2)(\rho^2+1)^{l-2}}{135\pi 2^{3(l+1)}\lambda_l\rho^{2l+1}}\times
~\nonumber~~~~~~~~~~~~~~~~~~~~~~~~~~~~~~~~~~~~~~~~~~~~~~~~~~~~~~~~~~~~~~~~~~~~~~~~~~~~~~~~~~~~~~~~~\\
\left[\rho^{2l+2}\sum_{n=0}^4 \mu_{ln}^{(1)}\rho^{n}+\rho\sum_{n=0}^l \mu_{ln}^{(2)}\rho^{2n}
-\mu_{l0}^{(2)}(\rho^2+1)^6 \arctan \rho \sum_{n=0}^{l-3}\mu_{ln}^{(3)}\rho^{2n}
\right]+A_2(l)v_{4l}(\rho),~~~~~~~(l\geq3)~~~~~
\end{eqnarray}
where $G$ is the Catalan's constant, and function $v_{4l}(\rho)$ is defined by Eq.(\ref{342b}) for $k=4$.
The $A_2(l)$-factors are presented by Eqs.(\ref{C40})-(\ref{C42}), whereas the
coefficients $\mu_{nl}^{(i)}~(i=1,2,3)$ and $\lambda_l$ can be found in Tables \ref{T2}, \ref{T3}.
It is important that $A_2(l)$ are equal to zero for odd $l$.
Note that for $l=1$ we have obtained the admixture coefficient $\mathcal{F}_{41}=0$ (see, Appendix \ref{SC}). This means that for all odd $l$ in expansion (\ref{718}), the solution of the form (\ref{340}) just gives the correct results for $\tau_l$ satisfying the coupling equation (\ref{546}) or the equivalent Eq.(\ref{C15}).

Remind that the general solution of the FRR (\ref{702}) must contain the addition $\sum_l a_{4l}Y_{4l}$, where the coefficients $a_{kl}$ can be determined only by analysis of the asymptotic behavior of the wave function (see, the end of Sec.\ref{S2}).
Using Eqs.(\ref{110})-(\ref{111}), one can detect the presence of admixture of the HH $Y_{4l}$ in any component $\psi_{4,1}^{(j)}$, and then get rid of such an admixture, just as it was done in Sec.\ref{S5b} for $\psi_{2,0}^{(1)}$. Therefore, only the "pure" components $\psi_{4,1}^{(j)}$ are presented by Eqs.(\ref{716})-(\ref{724}) and Table \ref{T1}.

The last subcomponent, we present here, is $\psi_{3,0}^{(2c)}$. It is the physical solution of Eq.(\ref{703}) with the rhs, $h_{3,0}^{(2c)}$ defined by Eq.(\ref{755}). To apply the technique described in Sec.\ref{S4}, we again use the single sum representation,
\begin{equation}\label{726}
\psi_{3,0}^{(2c)}(\alpha,\theta)=\sum_{l=0}^\infty P_l(\cos\theta)(\sin \alpha)^l \phi_l(\rho),
\end{equation}
where $\rho$ is defined by Eq.(\ref{334}).
Function $\phi_l(\rho)$ is obtained (details can be found in Appendix \ref{SD}) in the form:
\begin{equation}\label{727}
\phi_l(\rho)=\phi_l^{(p)}(\rho)+c_l v_{3l}(\rho),
\end{equation}
where the reduction of the formula (\ref{342b}) for $k=3$ and $0\leq\rho\leq1$ yields:
\begin{equation}\label{728}
v_{3l}(\rho)=\left(\rho^2+1\right)^{l-\frac{3}{2}}
\left[\frac{(2l-3)(2l-1)}{(2l+3)(2l+5)}\rho^4+\frac{2(2l-3)}{2l+3}\rho^2+1\right].
\end{equation}
The particular solution $\phi_l^{(p)}$ of the corresponding differential equation in $\rho$ (see, Eq.(\ref{D4})) can be presented in the form,
\begin{equation}\label{730}
\phi_l^{(p)}(\rho)=\frac{2^{-l}\left(\rho^2+1\right)^{l-\frac{3}{2}}}{3(2l-3)(2l-1)(2l+3)(2l+5)}
\left[2f_{1l}(\rho)+\frac{2f_{2l}(\rho)+f_{3l}(\rho)}{2l+1}\right],
\end{equation}
where
\begin{equation}\label{731}
f_{1l}(\rho)=\left[9-4l(l+2)\right]\rho+\left(13-4l^2\right)\rho^3,~~~~~~~~~~~~~~~~~~~~~~~~~~~~~~~~~~~~
\end{equation}
\begin{equation}\label{732}
f_{2l}(\rho)=\left[(2l-3)(2l-1)\rho^4+2(2l-3)(2l+5)\rho^2+(2l+3)(2l+5)\right]\arctan(\rho),~~~~~~~~~
\end{equation}
\begin{eqnarray}\label{733}
f_{3l}(\rho)=-\left[(2l+3)(2l+5)\rho^4+2(2l-3)(2l+5)\rho^2+(2l-3)(2l-1)\right]\times
\nonumber~~~~~~~~~~~~~~~~~~~~\\
\frac{\rho}{l+1}~_2F_1\left(1,l+1;l+2;-\rho^2\right).~~~~~~~~~~~~~~~~~
\end{eqnarray}
Note that the hypergeometric function presented in Eq.(\ref{733}) can be expressed through elementary functions (see, Eq.(\ref{D7a})).
The coefficient $c_l$ is defined by Eqs.(\ref{D16})-(\ref{D24}). It is clear that formulas (\ref{727})-(\ref{733}) are valid for $0\leq\alpha\leq\pi/2$.
For  $\pi/2<\alpha\leq\pi$, one should replace $\rho$ by $1/\rho$.

There are only two subcomponents, $\psi_{3,0}^{(1d)}$ and $\psi_{3,0}^{(2d)}$, that were not determined in this research.
The reason is that the right-hand sides $h_{3,0}^{(1d)}$ and $h_{3,0}^{(2d)}$ of the corresponding IFRR (see, Eqs.(\ref{754})-(\ref{755})) include the function $\chi_{20}$ presented by expansion (\ref{543}), what
complicates the calculations and dramatically increases the size of the final formulas.

\section{Conclusions} \label{S7}

Solutions of the Fock recurrence relations (\ref{103}) were used to derive the angular coefficients $\psi_{k,p}(\alpha,\theta)$ of the Fock expansion (\ref{1}) describing the $S$-state wave function of the two-electron atomic system. The hyperspherical coordinates  with hyperspherical angles (\ref{2}) were applied.

The FRR were separated into the independent individual equations (\ref{114}) associated with definite power $j$ ($p\leq j\leq k-p$) of the nucleus charge $Z$.
The appropriate solutions $\psi_{k,p}^{(j)}$ of Eq.(\ref{114})  present the independent components (of the AFC) defined by Eqs.(\ref{112})-(\ref{113}).

The property of finiteness at the boundary points of the hyperspherical angular space was extensively used for deriving each component.

The "pure" components not containing the admixture of the HH, $Y_{kl}$ are found for even values of $k$.

A few methods for solving the individual FRR were proposed. Simple technique for solving the IFRR with simplest nonseparable right-hand side (\ref{113}) was described in Sec.\ref{S3}, whereas effective method for solving the IFRR with separable right-hand side of a specific but frequent kind was presented in Sec.\ref{S4}.

Some mistakes/misprints made in references \cite{FOR}, \cite{LEZ1} for the closed analytic form of $\psi_{20}$,
and - in reference \cite{AM1} for the double and single infinite series representations of the component $\psi_{20}^{(1)}$, were noted and corrected.

The coupling equation (\ref{546}) was proposed and applied in case of a single series representation
for the component of the AFC.

Using the techniques mentioned above, all the components of the AFC, $\psi_{41}$ and the majority of components and subcomponents of $\psi_{30}$ were calculated and presented for the first time in Tables \ref{T1}-\ref{T3} and in explicit formulas of Sec.\ref{S6}.
Details of all the calculations have been placed in the Appendices \ref{SA}-\ref{SD}.

All kinds of calculations (both analytical and numerical) were carried out with the help of the program \emph{Mathematica} (Wolfram.com).
\section{Acknowledgment}
This work was supported by the Israel Science Foundation grant 954/09 and the "Pazy Foundation".
\appendix
\section{}\label{SA}
For ease of comparison, let us present $\psi_{2,0}^{(1)}$ defined by Eqs.(\ref{402}), (\ref{403b}) and (\ref{404}),  in the form
\begin{equation}\label{A1}
\psi_{2,0}^{(1)}=\chi_{20}(\alpha,\theta)+\varphi(\alpha,\theta),
\end{equation}
where
\begin{equation}\label{A2}
\varphi(\alpha,\theta)=-\frac{1}{3}\left[\sin(\alpha/2)+\cos(\alpha/2)\right]\sqrt{1-\sin \alpha \cos\theta}.
\end{equation}
By direct action of the HAM operator defined by Eq.(\ref{104}), one obtains:
\begin{equation}\label{A3}
\left(\Lambda^2-12\right)\varphi(\alpha,\theta)=h_{2,0}^{(1a)},
\end{equation}
where the rhs of Eq.(\ref{A3}) is defined by Eq.(\ref{404a}). Hence,
\begin{equation}\label{A4}
\left(\Lambda^2-12\right)\chi_{20}(\alpha,\theta)=h_{2,0}^{(1b)},
\end{equation}
where the rhs of Eq.(\ref{A4}) is defined by Eq.(\ref{404b}).

First of all, we need to solve Eq.(\ref{A4}) by expanding $\chi_{20}$ in HH. To this end, following \cite{AM1}, let us consider the HH expansion of the function
\begin{equation}\label{A5}
f(\alpha,\theta)\equiv\frac{\cos(\alpha/2)+\sin(\alpha/2)}{\sin\alpha\sqrt{1-\sin \alpha \cos\theta}}=\sum_{nl}D_{nl}Y_{nl}(\alpha,\theta),
\end{equation}
where $Y_{nl}$ are the unnormalized HH. It follows from (\ref{106}), (\ref{110}) and (\ref{111}) that
\begin{equation}\label{A6}
D_{nl}=N_{nl}^2\int f(\alpha,\theta) Y_{nl}(\alpha,\theta)d\Omega.
\end{equation}
According to the works \cite{AM1} and \cite{Sack}, the following representation holds for $\nu >-2$:
\begin{equation}\label{A7}
\xi^{\nu}=\frac{\sqrt{\pi}}{\Gamma(-\nu/2)}\sum_{l=0}^\infty P_l(\cos \theta)\frac{\Gamma(l-\nu/2)}{\Gamma(l+1/2)}\left(\frac{\sin\alpha}{2}\right)^l
F_{l,\nu}(\alpha),
\end{equation}
where
\begin{equation}\label{A9}
F_{l,\nu}(\alpha)=
~_2F_1\left(\frac{l}{2}-\frac{\nu}{4},\frac{l}{2}-\frac{\nu}{4}+\frac{1}{2};l+\frac{3}{2};\sin^2\alpha\right)
\end{equation}
presents the Gauss hypergeometric function, and $\xi$ is defined by Eq.(\ref{208}).

Inserting (\ref{106}) (for unnormalized HH) and (\ref{A7}) for $\nu=-1$  into (\ref{A6}), and using the orthogonality of the Legendre polynomials, one obtains
\begin{equation}\label{A8}
D_{nl}=N_{nl}^2\frac{\pi^2 2^{1-l}}{2l+1}\int_0^{\pi}(\sin \alpha)^{2l+1} C_{n/2-l}^{(l+1)}(\cos\alpha)f^+(\alpha)d\alpha,
\end{equation}
where
\begin{equation*}
f^+(\alpha)=\left[\cos(\alpha/2)+\sin(\alpha/2)\right]~_2F_1\left(\frac{l}{2}+\frac{1}{4},\frac{l}{2}+\frac{3}{4};l+\frac{3}{2};\sin^2\alpha\right).
\end{equation*}
It is important to emphasize the following substantial property. It may be verified that integral in the rhs of Eq.(\ref{A8}) differs from zero only for even values of $n/2-l$.

The following relation was proved to be correct (see, e.g., Eq.(90) \cite{AM1}):
\begin{equation}\label{A10}
f^+(\alpha)=~_2F_1\left(\frac{l}{2},\frac{l+1}{2};l+1;\sin^2\alpha\right)+
\frac{1}{2}\sin \alpha~_2F_1\left(\frac{l+l}{2},\frac{l}{2}+1;l+2;\sin^2\alpha\right).
\end{equation}
Inserting (\ref{A10}) into (\ref{A8}), one obtains:
\begin{equation}\label{A11}
D_{nl}=N_{nl}^2\frac{\pi^2 2^{1-l}}{2l+1}\left(\mathcal{I}_1+\frac{1}{2}\mathcal{I}_2\right),
\end{equation}
where
\begin{equation}\label{A12}
\mathcal{I}_1=\int_{-1}^{1}y^{2l} C_{n/2-l}^{(l+1)}(x)~_2F_1\left(\frac{l}{2},\frac{l+1}{2};l+1;y^2\right)dx,
\end{equation}
\begin{equation}\label{A13}
\mathcal{I}_2=\int_{-1}^{1}y^{2l+1} C_{n/2-l}^{(l+1)}(x)~_2F_1\left(\frac{l+1}{2},\frac{l}{2}+1;l+2;y^2\right)dx.
\end{equation}
We made the change of variable, corresponding to notations (\ref{241}), that is $x=\cos \alpha,~~y=\sin\alpha$. Remind the well-known formula of orthogonality for the Gegenbauer polynomials (in terms of $x,y$):
\begin{equation}\label{A14}
\int_{-1}^{1}y^{2l+1}C_m^{(l+1)}(x)C_n^{(l+1)}(x)dx=
\frac{\pi (n+2l+1)!}{2^{2l+1}n!(n+l+1)(l!)^2}\delta_{mn},
\end{equation}
where $\delta_{mn}$ is the Kronecker delta.
In order to apply Eq.(\ref{A14}) to reduction of integral (\ref{A12}), we propose to use expansion (\emph{A}9) from \cite{AM1} with $n=-1,~\nu=l+1$, what gives:
\begin{equation}\label{A15}
y^{-1}~_2F_1\left(\frac{l}{2},\frac{l+1}{2};l+1;y^2\right)=
\frac{(l!)^2}{\pi}\sum_{m=0}^\infty\frac{(l+2m+1)\Gamma^2\left(m+1/2\right)}{\Gamma^2(l+m+3/2)}
G_1(l,m)C_{2m}^{(l+1)}(x),~~~~~~~~~
\end{equation}
where
\begin{equation}\label{A16}
G_1(l,m)=~_3F_2
\left(\frac{l}{2},\frac{l+1}{2},\frac{1}{2};\frac{1}{2}-m,l+m+\frac{3}{2};1\right)
\end{equation}
is the generalized hypergeometric function.
Inserting (\ref{A15}) into (\ref{A12}), and using (\ref{A14}), one easily obtains:
\begin{equation}\label{A17}
\mathcal{I}_1=\frac{\Gamma(m+1/2)(l+m)!}{m!\Gamma(l+m+3/2)}G_1(l,m),
\end{equation}
where $m=n/4-l/2$.

In order to apply Eq.(\ref{A14}) to reduction of integral (\ref{A13}), we propose to use expansion (\emph{A}6) from \cite{AM1} with $\nu=l+1$, what gives:
\begin{equation}\label{A18}
~_2F_1\left(\frac{l+1}{2},\frac{l}{2}+1;l+2;y^2\right)=
\frac{(l+1)!}{\sqrt{\pi}}\sum_{m=0}^\infty\frac{(-1/4)^m\Gamma\left(m+1/2\right)}{(l+m+1)!},
G_2(l,m)C_{2m}^{(l+1)}(x),~~~~~~~~~
\end{equation}
where
\begin{equation}\label{A19}
G_2(l,m)=~_3F_2
\left(\frac{l}{2}+m+\frac{1}{2},\frac{l}{2}+m+1,l+m+\frac{3}{2};l+m+2,l+2m+2;1\right).
\end{equation}
Inserting (\ref{A19}) into (\ref{A13}), and using (\ref{A14}), one obtains:
\begin{equation}\label{A20}
\mathcal{I}_2=\frac{\pi(l+1)(-1)^m(2l+2m+1)!}{2^{n+1}l!m!(l+2m+1)(l+m+1)!}G_2(l,m),
\end{equation}
where again $m=n/4-l/2$.
Substitution of Eqs.(\ref{A17}), (\ref{A20}) and (\ref{107}) into (\ref{A11}) yields finally:
\begin{eqnarray}\label{A21}
D_{nl}=\frac{2^l(l!)^2(2m)!}{\sqrt{\pi}m!(2l+2m+1)!}
\left[\frac{2(l+2m+1)\Gamma(m+1/2)(l+m)!}{\sqrt{\pi}\Gamma(l+m+3/2)}G_1(l,m)+
\right.
\nonumber~~~~~~~~~~~~~~~~~~~\\
\left.
\frac{(l+1)(-1)^m\Gamma(l+m+3/2)}{2^{2m}l!(l+m+1)}G_2(l,m)\right],~~~~~~~~~~~~~~~
\end{eqnarray}
where $m=n/4-l/2$ is a non-negative integer.
By direct numerical comparing with definition (\ref{A6}),
it is easy to make sure that relation (\ref{A21}) is correct, whereas formula (93) in \cite{AM1} is not correct.
Note that Eq.(\ref{A21}) is not a single representation for $D_{nl}$. We have found, at least, two another representations.

It should be emphasized that representation (\ref{A21}) for $D_{nl}$ is correct only for even $n/2-l$, otherwise, $D_{nl}=0$.
In particular, $D_{20}=0$. It follows from (\ref{A21}) that $D_{21}=4-8/\pi$, what means that using Eq.(\ref{A5}) and Eq.(\ref{404b}), one can rewrite Eq.(\ref{A4}) in the form
\begin{equation}\label{A22}
\left(\Lambda^2-12\right)\chi_{20}(\alpha,\theta)=\frac{2}{3}{\sum_{nl}}'D_{nl}Y_{nl}(\alpha,\theta),
\end{equation}
where the prime indicates that $n=2$ is omitted from the summation.
Taking into account that eigenvalues of the $\Lambda^2$-operator are given by $n(n+4)$,
one obtains for HH expansion, $\sum_{nl}X_{nl}Y_{nl}$ of the function $\chi_{20}$
\begin{equation}\label{A23}
\left(\Lambda^2-12\right)\chi_{20}(\alpha,\theta)={\sum_{nl}}'X_{nl}(n-2)(n+6)Y_{nl}(\alpha,\theta).
\end{equation}
Comparison of (\ref{A22}) and (\ref{A23}) yields
\begin{equation}\label{A24}
\chi_{20}(\alpha,\theta)=\frac{2}{3}{\sum_{nl}}'\frac{D_{nl}}{(n-2)(n+6)}Y_{nl}(\alpha,\theta).
\end{equation}
Remind, that contribution of the terms with $n=2$ must be treated separately as the solutions of the associated homogeneous equation.

\section{}\label{SB}

To obtain the analytical representations for $\sigma_l$, it is necessary to substitute expansion (\ref{543}) and representation (\ref{A7}) for $\nu=-1$ into the lhs and rhs of Eq.(\ref{A4}), respectively. Subsequent application of the relations (\ref{330})-(\ref{331}) for $k=2$ yields
\begin{eqnarray}\label{B1}
\sum_{l=0}^\infty P_l(\cos \theta)(\sin \alpha)^l\left[4\sigma_l''(\alpha)+8(l+1)\cot \alpha~\sigma_l'(\alpha)-4(l-1)(l+3)\sigma_l(\alpha)\right]=
\frac{8(\pi-2)}{3\pi} \sin \alpha \cos\theta
\nonumber~\\
-\frac{1}{3}\left[\csc\left(\frac{\alpha}{2}\right)+\sec\left(\frac{\alpha}{2}\right)\right]
\sum_{l=0}^\infty P_l(\cos \theta)\left(\frac{\sin \alpha}{2}\right)^l
~_2F_1\left(\frac{l}{2}+\frac{1}{4},\frac{l}{2}+\frac{3}{4};l+\frac{3}{2};\sin^2 \alpha\right).~~~~~~~~~~
\end{eqnarray}
Equating the expansion coefficients for the Legendre polynomials (in $\cos \theta$) of the same order, one obtains the ordinary differential equations that must be solved using the boundary conditions \textbf{P2} and the coupling equation (\ref{547}).
We shall solve these equations using the $\rho$ variable defined by Eq.(\ref{334}). Given that
\begin{equation}\label{B2}
\sin \alpha=2\rho/(\rho^2+1),~~~~~~~\csc(\alpha/2)+\sec(\alpha/2)=\rho^{-1}(\rho+1)\sqrt{\rho^2+1},~~~~~~~~~~
\end{equation}
the following relationship will be useful for further consideration:
\begin{equation}\label{B3}
~_2F_1\left(\frac{l}{2}+\frac{1}{4},\frac{l}{2}+\frac{3}{4};l+\frac{3}{2};\frac{4\rho^2}{(\rho^2+1)^2}\right)=
\left\{ \begin{array}{c}~~\mathlarger{(\rho^2+1)^{l+\frac{1}{2}} },~~~~~~~~~~~~~0\leq\rho\leq1 \\
\mathlarger{(\rho^2+1)^{l+\frac{1}{2}}\rho^{-2l-1} }.~~~~~~~~\rho>1
\end{array}\right.
\end{equation}

Special cases of $l=0,~l=1$ and $l\geq2$ will be considered.

\subsection{$\boldsymbol{l=0}$}\label{SB1}

Equating coefficients for the Legendre polynomials of zero order ($l=0$) in both sides of Eq.(\ref{B1}), and turning to the $\rho$ variable,
one can employ Eq.(\ref{335}) for $k=2$ and $l=0$. This yields
\begin{equation}\label{B4}
(\rho^2+1)^2S_0''(\rho)+\frac{2(\rho^2+1)}{\rho}S_0'(\rho)+12S_0(\rho)=-h(\rho),
\end{equation}
where the substitution $g(\rho)=S_0(\rho)\equiv \sigma_0(\alpha)$ was applied.
Using Eqs.(\ref{B2}), (\ref{B3}), one obtains for the rhs in the range $\rho\in[0,1]~~(0\leq\alpha\leq\pi/2)$:
\begin{equation}\label{B5}
h(\rho)=\frac{(\rho+1)(\rho^2+1)}{3\rho}.~~~~~~~~~~~~~~~
\end{equation}
For $l=0$ one can employ formula (\ref{340}) with $k=2,~\rho_c=1$, that gives the following solution of Eq.(\ref{B4})
for $0\leq\rho\leq1$:
\begin{equation}\label{B6}
S_0(\rho)=\frac{\rho\left\{2(\rho^2-1)\ln[(\rho^2+1)/2]-\rho(3\rho+2)-1\right\}-(\rho^4-6\rho^2+1)\arctan \rho}{12\rho(\rho^2+1)}.
\end{equation}
An alternative representation is:
\begin{equation}\label{B7}
\sigma_0=\frac{1}{12}\left\{\left(2y-\frac{1}{y}\right)\alpha+x
\left[1+ 2\ln(x+1) \right]-y-2 \right\},~~~~~~~~~~~0\leq\alpha\leq\pi/2
\end{equation}
where $x$ and $y$ are defined by Eq.(\ref{241}).
It is clear that for $\pi/2<\alpha\leq\pi$, one needs to replace $\rho$ by $1/\rho$, $x$ by $-x$, and $\alpha$ by $\pi-\alpha$. Representation (\ref{B7}) is simpler than the corresponding one (\emph{A}19) from \cite{AM1} which meanwhile is correct only for $0\leq\alpha\leq\pi/2$.

It is easy to verify that $\sigma_0$ presented by Eq.(\ref{B7}) corresponds to the "pure" component $\psi_{20}^{(1)}$
(see, the end of Sec.\ref{S1}), because the HH expansion coefficient
\begin{equation*}
\mathcal{F}_{20}\propto\int \sigma_0(\alpha)Y_{20}(\alpha,\theta)d\Omega
\end{equation*}
equals zero.

\subsection{$\boldsymbol{l=1}$}\label{SB2}

In this case, equating coefficients for $P_1(\cos \theta)=\cos \theta$ in both sides of Eq.(\ref{B1}), using (\ref{B2}), (\ref{B3}) and simplifying, one obtains the equation
\begin{equation}\label{B8}
(1+\rho^2)^2 S_1''(\rho)+\frac{2(2-\rho^2)(1+\rho^2)}{\rho}S_1'(\rho)=\frac{8(\pi-2)}{3\pi}-\frac{(1+\rho)(1+\rho^2)^2}{6\rho},
~~~~~~~~~~~0\leq\rho\leq 1
\end{equation}
where $S_1(\rho)\equiv\sigma_1(\alpha)$.
Formula (\ref{340}) with $k=2,~l=1,~\rho_c=1$ yields the particular solution to Eq.(\ref{B8}) in the form
\begin{eqnarray}\label{B10}
S_1^{(p)}(\rho)=\frac{1}{24\pi}\left[\left(\frac{1}{\rho^3}+\frac{9}{\rho}-9\rho-\rho^3\right)\arctan \rho -
\frac{1}{\rho^2}-\rho^2+2-\frac{2\pi}{3}\right]-
\nonumber~~~~~~~~~~~~~~~~~~~\\
\frac{\rho\left(\rho^2-6\rho+3\right)}{72}+
\frac{1}{6}\ln\left(\frac{1+\rho^2}{2}\right).~~~~~~~
\end{eqnarray}
To obtain the "pure" (single-valued) solution to Eq.(\ref{B8}), one needs to calculate the coefficient $\mathcal{F}_{21}$ of the unnormalized HH expansion of the solution (\ref{B10}).
Remind that representation (\ref{B10}) is correct only for $0\leq\rho\leq1~(0\leq\alpha\leq\pi/2)$.
For $\rho>1~(\pi/2<\alpha\leq\pi)$, one should replace $\rho$ by $1/\rho$ in the rhs of Eq.(\ref{B10}).
Thus, using Eqs.(\ref{110})-(\ref{111}) for $n=2,~l=1$ one obtains:
\begin{eqnarray}\label{B11}
\mathcal{F}_{21}=N_{21}^2\int \sigma_1^{(p)}(\alpha)Y_{21}(\alpha,\vartheta)d\Omega=
64N_{21}^2\pi^2\int_0^1 \frac{S_1^{(p)}(\rho)\rho^4}{(1+\rho^2)^5}d\rho \int_0^{\pi}\sin \theta \cos^2\theta d\theta=
\nonumber~~~~~~~~~~~~~\\
\frac{14-48 G+\pi(1+12 \ln 2)}{72 \pi},~~~~~~~~~~~~~~~~~~~~~~~~~~~~~~~~~~~~~~~~~~~~~~~~~~~~~~~~~~~
\end{eqnarray}
where $G$ is the Catalan's constant, and $N_{21}$ is defined by Eq.(\ref{107}).
The final result is: $\sigma_1=S_1^{(p)}(\rho)-\mathcal{F}_{21}$.

The same result can be obtained using Eq.(\ref{547}), which for $l=1$ reduces to
\begin{equation}\label{B13}
\int_0^{\pi/2} y^4 \sigma_1(\alpha)d\alpha=0,
\end{equation}
where $\sigma_1(\alpha)$ presents the general solution of Eq.(\ref{B8}) satisfying the condition \textbf{P2} of finiteness.

\subsection{$\boldsymbol{l\geq 2}$}\label{SB3}

The required function $\sigma_l~(l\geq2)$ can be certainly obtained by means of application of the particular solution (\ref{340}), and subsequent use of Eq.(\ref{547}). However, in this case we shall apply the following simplified procedure.

Equating coefficients for $P_l(\cos \theta)$ in both sides of Eq.(\ref{B1}) (for $l\geq 2$), using (\ref{B2}), (\ref{B3}), and introducing the function $Q_l(\rho)$ by means of the relationship
\begin{equation}\label{B17}
\sigma_l(\alpha)\equiv S_{l}(\rho)=-\frac{2^{-l}}{3}(\rho^2+1)^{l-1}Q_l(\rho),
\end{equation}
one obtains the differential equation
\begin{equation}\label{B18}
\rho(\rho^2+1)Q_l''(\rho)+2\left[(l-2)\rho^2+l+1\right]Q_l'(\varrho)-6(l-1)\rho~Q_l(\rho)=(\rho+1)(\rho^2+1),
\end{equation}
which is correct only for $0\leq\rho\leq1~(0\leq\alpha\leq\pi/2)$. \emph{Mathematica} gives the following general solution to the homogeneous equation associated with inhomogeneous equation (\ref{B18}):
\begin{equation}\label{B19}
Q_l^{(h)}(\rho)=C_1~_2F_1\left(-\frac{3}{2},l-1;l+\frac{3}{2}; -\rho^2\right)+
C_2\rho^{-2l-1}~_2F_1\left(-\frac{3}{2},-l-2;\frac{1}{2}-l; -\rho^2\right).
\end{equation}
Assuming that a particular solution has a form $a+b\rho+c\rho^2+d\rho^3$, and substituting the latter one into the lhs of Eq.(\ref{B18}), one easily find this particular solution in the form:
\begin{equation}\label{B20}
Q_l^{(p)}(\rho)=\frac{1}{2}\left[\frac{l\rho^3}{(l+1)(l+2)}-\frac{\rho^2}{l}+\frac{\rho}{l+1}-\frac{l+1}{l(l-1)}\right].
\end{equation}
It is seen that one should put $C_2=0$ to satisfy the condition \textbf{P2} of finiteness.
Thus, using (\ref{B17}), (\ref{B19}) and (\ref{B20}) we presently obtain:
\begin{eqnarray}\label{B21}
S_l(\rho)=-\frac{2^{-l-1}}{3}\left\{(1+\rho^2)^{l-1}\left[\frac{l\rho^3}{(l+1)(l+2)}-\frac{\rho^2}{l}+\frac{\rho}{l+1}-\frac{l+1}{l(l-1)}\right]+
\right.
\nonumber~~~~~~~~~~~~~~~\\
\left.
2C_1~_2F_1\left(\frac{l-1}{2},\frac{l+3}{2};l+\frac{3}{2};y^2\right)\right\},~~~~~~~~~~~
\end{eqnarray}
where the relationship
\begin{equation*}
~_2F_1\left(\frac{l-1}{2},\frac{l+3}{2};l+\frac{3}{2};y^2\right)=(1+\rho^2)^{l-1}~_2F_1\left(-\frac{3}{2},l-1;l+\frac{3}{2};-\rho^2 \right)
\end{equation*}
was used (see, 7.3.1.54 \cite{PRU3}).
Now we can apply the coupling equation (\ref{547}) to find the constant $C_1\equiv C_1(l)$. To this end, first, we need to calculate explicitly the integral
\begin{equation}\label{B22}
\int_0^\pi y^{2l+2}\sigma_l(\alpha)d\alpha=\mathcal{K}_1(l)+C_1\mathcal{K}_2(l),
\end{equation}
where, using Eq.(\ref{B21}) one obtains:
\begin{equation}\label{B23}
\mathcal{K}_1(l)=-\frac{2^{l+3}}{3}\int_0^1\frac{\rho^{2l+2}}{(\rho^2+1)^{l+4}}
\left[\frac{l\rho^3}{(l+1)(l+2)}-\frac{\rho^2}{l}+\frac{\rho}{l+1}-\frac{l+1}{l(l-1)}\right]d\rho,~~~~~
\end{equation}
\begin{equation}\label{B24}
\mathcal{K}_2(l)=-\frac{2^{-l}}{3}\int_{-1}^1 y^{2l+1}
~_2F_1\left(\frac{l-1}{2},\frac{l+3}{2};l+\frac{3}{2};y^2 \right) d x .~~~~~~~~~~~~~~~~~~~~~~~~~~~~~~~~
\end{equation}
\emph{Mathematica} gives for the integral (\ref{B23}):
\begin{eqnarray}\label{B25}
\mathcal{K}_1(l)=\frac{1}{3l}\left[\frac{l+1}{(l-1)(2l+3)}
~_2F_1\left(-\frac{3}{2},1;l+\frac{5}{2};-1 \right)+\frac{1}{2l+5}~_2F_1\left(-\frac{1}{2},1;l+\frac{7}{2};-1 \right)-
\right.
\nonumber~~~~~~~\\
\left.
\frac{l}{(l+1)(l+3)}
\right].~~~~~~~~~~~~~
\end{eqnarray}
To reduce integral (\ref{B24}) we applied the expansion (\emph{A}9) from \cite{AM1}, which for parametrization $n=l,~\nu=1/2$ becomes:
\begin{eqnarray*}
y^{2l+1}~_2F_1\left(\frac{l-1}{2},\frac{l+3}{2};l+\frac{3}{2};y^2 \right)=
\frac{1}{2}\Gamma^2\left(l+\frac{3}{2}\right)\sum_{m=0}^\infty \frac{(-1)^m(4m+1)\Gamma\left(m+1/2\right)}{m!(l+m+1)!\Gamma\left(l-m+3/2\right)}\times
\nonumber~~~~~~~~~~~~\\
~_3F_2\left(\frac{l-1}{2},\frac{l+3}{2},l+\frac{3}{2};l+\frac{3}{2}-m,l+m+2;1 \right)C_{2m}^{(1/2)}(x).~~~~~~~~~~
\end{eqnarray*}
Substitution of the latter representation into Eq.(\ref{B24}), and subsequent application of the formula of orthogonality for the Gegenbauer polynomials, yields finally:
\begin{equation}\label{B26}
\mathcal{K}_2(l)=-\frac{2^{-l}\sqrt{\pi} \Gamma\left(l+\frac{3}{2}\right)}{3\Gamma\left(\frac{l+1}{2}\right)\Gamma\left(\frac{l+5}{2}\right)}.~~~~~~~~~~
\end{equation}
Using Eq.(\ref{547}) and Eq.(\ref{B22}), one obtains:
\begin{equation}\label{B27}
C_1=\frac{\mathcal{K}(l)}{3(l-1)(l+1)(l+3)\mathcal{K}_2(l)},
\end{equation}
where
\begin{equation}\label{B28}
\mathcal{K}(l)=\mathcal{K}_3(l)+1-3(l-1)(l+1)(l+3)\mathcal{K}_1(l),
\end{equation}
\begin{equation}\label{B29}
\mathcal{K}_3(l)=\frac{\sqrt{\pi} \Gamma(l+3/2)}{2^{l+2}l!}~_3F_2\left(\frac{l+1}{2},\frac{l}{2}+1,l+\frac{3}{2};l+2,l+2;1\right).
\end{equation}
The simplest way for reduction of $\mathcal{K}(l)$ is to calculate numerically the first entries of the sequence, that is  to calculate $\mathcal{K}(l)$ for $l=2,3,4,5$. Using \emph{Mathematica}, one obtains the required sequence: $-5/2,-7/3,-9/4,-11/5$. It is seen that $\mathcal{K}(l)=-(2l+1)/l$. Simplifying, we have, finally:
\begin{equation}\label{B30}
C_1=\frac{(l-2)!\Gamma\left(\frac{l+1}{2}\right)}{2\Gamma\left(l+\frac{1}{2}\right)\Gamma\left(\frac{l}{2}+1\right)}.
\end{equation}

\section{}\label{SC}

Solution of the FRR (\ref{704}) for subcomponent with $j=2d$ implies that we need to find a suitable solution to the equation
\begin{equation}\label{C1}
\left(\Lambda^2-32\right)\psi_{4,1}^{(2d)}=h_{4,1}^{(2d)},~~~~~~~~~~~~~~~~~
\end{equation}
where $h_{4,1}^{(2d)}$ and $\psi_{4,1}^{(2d)}$ are defined by Eq.(\ref{756b}) and expansion (\ref{718}), respectively.
Inserting expansion (\ref{A7}) for $\nu=-1,1,3$ into the rhs of Eq.(\ref{756b}), one obtains the following expansion
\begin{equation}\label{C2}
h_{4,1}^{(2d)}(\alpha,\theta)=\sum_{l=0}^\infty P_l(\cos\theta)(\sin \alpha)^l \textmd{h}_l(\alpha),
\end{equation}
for the rhs of Eq.(\ref{C1}), where
\begin{eqnarray}\label{C3}
\textmd{h}_l(\alpha)=\frac{\pi-2}{3\pi(2l-1)2^l}
\left[\sin\left(\frac{\alpha}{2}\right)+\cos\left(\frac{\alpha}{2}\right) \right]\times
\nonumber~~~~~~~~~~~~~~~~~~~~~~~~~~~~~~~~~~~~~~~~~~~~~~\\
\left[\frac{5}{(2l-3)\sin \alpha}F_{l,3}(\alpha)-\left(1-\frac{2}{\sin\alpha}\right)F_{l,1}(\alpha)-
(2l-1)F_{l,-1}(\alpha)\right]
,~~~~~~~~~~~
\end{eqnarray}
and where the hypergeometric function $F_{l,\nu}$ is defined by Eq.(\ref{A9}).
Thus, according to Eq.(\ref{331}) with $\textmd{g}(\alpha)=\textmd{t}_l(\alpha)$ and $k=4$, and expansions (\ref{718}) and (\ref{C2}), one obtains the following differential equation
\begin{equation}\label{C4}
4\textmd{t}_l''(\alpha)+8(l+1)\cot \alpha~\textmd{t}_l'(\alpha)+4(2-l)(l+4)\textmd{t}_l(\alpha)=-\textmd{h}_l(\alpha)
\end{equation}
for the function $\textmd{t}_l(\alpha)$ defined by Eq.(\ref{718}).
Transformation of Eq.(\ref{C4}) to variable $\rho=\tan(\alpha/2)$ yields (see, Eq.(\ref{335}))
\begin{equation}\label{C5}
\left(1+\rho^2\right)^2\tau_l''(\rho)+2\rho^{-1}\left[1+\rho^2+l(1-\rho^4)\right]\tau_l'(\rho)+4(2-l)(l+4)\tau_l(\rho)=
-h_l(\rho),
\end{equation}
where $h_l(\rho)\equiv \textmd{h}_l(\alpha)$, and $\tau_l(\rho)\equiv \textmd{t}_l(\alpha) $. Turning to the variable $\rho$ in Eq.(\ref{C3}), and simplifying, one obtains the following compact representation for the rhs of Eq.(C5):
\begin{eqnarray}\label{C6}
h_l(\rho)=-\frac{(\pi-2)(\rho+1)\left(\rho^2+1\right)^{l-1}}{3\pi(2l-1)(2l+3)2^{l+1}}\times
\nonumber~~~~~~~~~~~~~~~~~~~~~~~~~~~~~~~~~~~~~~~~~~~~~~~~~\\
\left[\frac{15-4l(l+1)(4l+11)}{(2l-3)(2l+5)\rho}+4l(2l+3)+2\rho+4(l+1)(2l-1)\rho^2+\frac{(2l-1)(4l+5)\rho^3}{2l+5}\right],~~~
\end{eqnarray}
where $0\leq\rho\leq1$. One of the possible ways to solve Eq.(\ref{C5}) is to take into account that its rhs (\ref{C6}) presents some linear combination of the terms of the form:
\begin{equation}\label{C7}
h_{n,l}(\rho)=\rho^n(\rho+1)(\rho^2+1)^{l-1},
\end{equation}
with integer $n>-2$. The particular solution, $T_{n,l}(\rho)$ of Eq.(\ref{C5}) with changing $h_{l}(\rho)$ for $h_{n,l}(\rho)$ defined by Eq.(\ref{C7}), reads:
\begin{eqnarray}\label{C8}
T_{n,l}(\rho)=\frac{\left(\rho^2+1\right)^{l+4}\rho^{n+2}}{2l+1}
\left\{~_2F_1\left(\frac{7}{2},3-l;\frac{1}{2}-l;-\rho^2\right)\times
\right.
\nonumber~~~~~~~~~~~~~~~~~~~~~~~~~~~~~~~~\\
\left[\frac{~_3F_2\left(\frac{7}{2},l+4,l+\frac{n+4}{2};l+\frac{3}{2},l+\frac{n+6}{2};-\rho^2\right)\rho}
{2l+n+4}+
\frac{~_3F_2\left(\frac{7}{2},l+4,l+\frac{n+3}{2};l+\frac{3}{2},l+\frac{n+5}{2};-\rho^2\right)}
{2l+n+3}
\right]
\nonumber~\\
-~_2F_1\left(\frac{7}{2},l+4;l+\frac{3}{2};-\rho^2\right)\times
\nonumber~~~~~~~~~~~~~~~~~~~~~~~~~~~~~~~~\\
\left.
\left[\frac{~_3F_2\left(\frac{7}{2},3-l,\frac{n+3}{2};\frac{1}{2}-l,\frac{n+5}{2};-\rho^2\right)\rho}
{n+3}+
\frac{~_3F_2\left(\frac{7}{2},3-l,\frac{n}{2}+1;\frac{1}{2}-l,\frac{n}{2}+2;-\rho^2\right)}
{n+2}
\right]\right\}.~~~~~~~
\end{eqnarray}
This solution was derived by means of application of Eq.(\ref{340}) with changing definite integration for indefinite integration over $\rho$ (that is putting the antiderivatives for the lower limit equal zero).

Here, we present the particular solutions, $\tau_l^{(p)}$ of Eq.(\ref{C5}) obtained by the exact formula (\ref{340}) with $k=4$. We consider the cases of $l=0,1,2$ (for $l_{max}=k/2$), and $l\geq3$, separately.

Application of Eqs.(\ref{340})-(\ref{343}) with $h(\rho)=h_l(\rho)$ defined by Eq.(\ref{C6}), and $\rho_c=1$ yields for $l=0,1,2$:
\begin{eqnarray}\label{C9}
\tau_{0}^{(p)}(\rho)=-\frac{\pi-2}{108\pi\left(\rho^2+1\right)^2}
\left[\frac{(1-\rho)}{15}\left(19\rho^4+142\rho^3+82\rho^2-48\rho-3\right)-
\right.
\nonumber~~~~~~~~~~~~~~~~~~~~~~~~~~~~~~~~\\
\left.
\left(\rho^5-15\rho^3+15\rho-\frac{1}{\rho}\right)\arctan \rho+\left(3\rho^4-10\rho^2+3\right)\ln\left(\frac{\rho^2+1}{2}\right)
\right],~~~~~~~
\end{eqnarray}
\begin{eqnarray}\label{C10}
\tau_{1}^{(p)}(\rho)=\frac{\pi-2}{302400\pi\rho^2\left(\rho^2+1\right)}
\left[1268\rho^7-2505\rho^6+1960\rho^5+32263\rho^4+18900\rho^3+18305\rho^2-735+
\right.
\nonumber~\\
\left.
735\left(\rho^7+20\rho^5-90\rho^3+20\rho+\frac{1}{\rho}\right)\arctan \rho-
23520\rho^2(\rho^2-1)\ln\left(\frac{\rho^2+1}{2}\right)
\right],~~~~~~~~~~~~~~~~~
\end{eqnarray}
\begin{eqnarray}\label{C11}
\tau_{2}^{(p)}(\rho)=\frac{\pi-2}{362880\pi\rho^4}
\left[672\rho^9-465\rho^8+760\rho^7-6720\rho^6-5880\rho^5+8798\rho^4+2520\rho^2+315+
\right.
\nonumber~\\
\left.
105(\rho^2-1)\left(3\rho^7+28\rho^5+178\rho^3+28\rho+\frac{3}{\rho}\right)\arctan \rho-
13440\rho^4\ln\left(\frac{\rho^2+1}{2}\right)
\right].~~~~~~~~~
\end{eqnarray}
It was mentioned in Sec.\ref{S1} that in order to obtain the "pure" component $\psi_{4,1}^{(2d)}$ one needs to select and then to get rid of admixture of the  HH, $Y_{4l}$.
In other words, we need to orthogonalize  $\psi_{4,1}^{(2d)}$ to each of  $Y_{4l}$.
First of all, according to expansion (\ref{718}) and Eqs.(\ref{110})-(\ref{111}) we need to calculate the unnormalized HH expansion coefficients
\begin{eqnarray}\label{C12}
\mathcal{F}_{4l}=\pi^2N_{4l}^2\left[\int_0^{\pi/2}d\alpha~ \tau_l^{(p)}(\rho)(\sin\alpha)^{l+2}
\int_0^\pi Y_{4l}(\alpha,\theta)P_l(\cos \theta)\sin \theta d \theta+
\right.
\nonumber~~~~~~~~~~~~~~~~~~~~\\
\left.
\int_{\pi/2}^{\pi}d\alpha~ \tau_l^{(p)}(1/\rho)(\sin\alpha)^{l+2}
\int_0^\pi Y_{4l}(\alpha,\theta)P_l(\cos \theta)\sin \theta d \theta
\right],~~~~~~~~
\end{eqnarray}
where the factor $N_{4l}$ is defined by Eq.(\ref{107}). The direct integration with the use of solutions (\ref{C9})-(\ref{C11}) yields:
\begin{subequations}\label{C13}
\begin{align}
\mathcal{F}_{40}=-\frac{(\pi-2)}{8100\pi^2}\left[5\pi(15\ln 2-16)+247-300 G \right],~~~~~~~~~~~~~~~~~~~~~~~~~\label{C13a}\\
\mathcal{F}_{41}=0,~~~~~~~~~~~~~~~~~~~~~~~~~~~~~~~~~~~~~~~~~~~~~~~~~~~~~~~~~~~~~~~~~~~~~~~~~~~\label{C13b}\\
\mathcal{F}_{42}=-\frac{(\pi-2)}{113400\pi^2}\left[5\pi(840\ln 2+109)+4592-16800 G \right],~~~~~~~~~~~~~~\label{C13c}
\end{align}
\end{subequations}
where $G$ is the Catalan's constant. Thus, to obtain the "pure" subcomponent $\psi_{4,1}^{(2d)}$, one needs
to substitute $\tau_0^{(p)}$ by $\tau_0=\tau_0^{(p)}-\mathcal{F}_{40}(4\cos^2\alpha-1)$, and $ \tau_2^{(p)}$ by $\tau_2=\tau_2^{(p)}-\mathcal{F}_{42}$ in expansion (\ref{718}).

For $l\geq3$, the particular solution of Eq.(\ref{C5}) obtained by Eq.(\ref{340}), can be presented in the form:
\begin{equation}\label{C14}
\tau_{l}^{(p)}(\rho)=\frac{(\pi-2)(\rho^2+1)^{l-2}}{135\pi 2^{3(l+1)}\lambda_l\rho^{2l+1}}
\left[\rho^{2l+2}\sum_{n=0}^4 \mu_{ln}^{(1)}\rho^{n}+\rho\sum_{n=0}^l \mu_{ln}^{(2)}\rho^{2n}
-\mu_{l0}^{(2)}(\rho^2+1)^6 \arctan \rho \sum_{n=0}^{l-3}\mu_{ln}^{(3)}\rho^{2n}
\right],
\end{equation}
where the coefficients $\lambda_l$ and $\mu_{ln}^{(1)}$ are presented in Table \ref{T2}, whereas $\mu_{ln}^{(2)}$ and $\mu_{ln}^{(3)}$ are presented in Table \ref{T3} in the form of lists. The number of coefficients is limited by $ l\leq10$.

It is clear that, all the terms of expansion (\ref{718}) with $l\geq3$ are automatically orthogonal to the $Y_{4l}$.
Therefore, in this case we shall use relation (\ref{546}), which becomes:
\begin{equation}\label{C15}
\mathcal{T}_{2l,l}=\frac{(l+1)!}{\sqrt{\pi}\Gamma(l+3/2)}\int_0^{\pi} \textmd{t}_l(\alpha)(\sin \alpha)^{2l+2}d\alpha=
\frac{2^{2(l+2)}(l+1)!}{\sqrt{\pi}\Gamma(l+3/2)}\int_0^{1} \tau_l(\rho)\frac{\rho^{2l+2}}{(1+\rho^2)^{2l+3}}d\rho,~~~~~~~~
\end{equation}
where $\mathcal{T}_{n,l}$ are the unnormalized HH expansion coefficients for the considered subcomponent
\begin{equation}\label{C16}
\psi_{4,1}^{(2d)}(\alpha,\theta)={\sum_{nl}}'\mathcal{T}_{n,l}Y_{nl}(\alpha,\theta).
\end{equation}
The prime indicates that $n=4$ is omitted from the summation for "pure" component.

The general solution of Eq.(\ref{C5}) reads
\begin{equation}\label{C17}
\tau_l(\rho)=\tau_l^{(p)}(\rho)+A_1 u_{4l}(\rho)+A_2 v_{4l}(\rho),
\end{equation}
where the solutions $u_{4l}$ and $v_{4l}$ of the homogeneous equation associated with Eq.(\ref{C5}) are defined by Eqs.(\ref{342}), and the constants $A_1,~A_2$ are currently undetermined.
It is clear that in order $\psi_{4,1}^{(2d)}$ satisfies the finiteness condition \textbf{P2}, function $(\sin \alpha)^l \tau_l(\rho)$ for each $l$ must satisfy this condition, according to expansion (\ref{718}).
It is easy to verify that $(\sin \alpha)^l u_{4l}(\rho)$ is divergent at $\rho=0$ ($\alpha\rightarrow 0$), whereas
both functions $(\sin \alpha)^l v_{4l}(\rho)$ and $(\sin \alpha)^l \tau_l^{(p)}(\rho)$ are finite at this point.
Hence, one should put $A_1=0$. Thus, Eq.(\ref{C15}) can be rewritten in the form:
\begin{equation}\label{C20}
\mathcal{T}_{2l,l}=
\frac{2^{2(l+2)}(l+1)!}{\sqrt{\pi}\Gamma(l+3/2)}
\left[\mathcal{P}_{1}(l)+A_2\mathcal{P}_{2}(l) \right],~~~~~~~~
\end{equation}
where
\begin{equation}\label{C21}
\mathcal{P}_{1}(l)=
\int_0^{1} \tau_l^{(p)}(\rho)\frac{\rho^{2l+2}}{(1+\rho^2)^{2l+3}}d\rho,~~~~~~~~
\end{equation}
\begin{equation}\label{C22}
\mathcal{P}_{2}(l)=
\int_0^{1} v_{4l}(\rho)\frac{\rho^{2l+2}}{(1+\rho^2)^{2l+3}}d\rho=
\frac{\sqrt{\pi}~2^{-2(l+2)}\Gamma\left(l+3/2\right)}{\Gamma(l/2+3)\Gamma(l/2)}.~~~~~~~~
\end{equation}
Note that for each value of $l\geq3$, the integral (\ref{C21}) can be calculated in the exact explicit form.

To derive the closed expression for $\mathcal{T}_{n,l}$ according to definition (\ref{C16}), first, let us obtain the HH expansion for the rhs of Eq.(\ref{C1}),
\begin{equation}\label{C25}
h_{4,1}^{(2d)}=\sum_{nl}\mathcal{H}_{n,l}^{(4)}Y_{nl}(\alpha,\theta),
\end{equation}
where according to Eq.(\ref{111}) for unnormalized HH,
\begin{equation}\label{C28}
\mathcal{H}_{n,l}^{(4)}=N_{nl}^2\int h_{4,1}^{(2d)}(\alpha,\theta)Y_{nl}(\alpha,\theta)d\Omega,
\end{equation}
and the normalization coefficient $N_{nl}$ is defined by Eq.(\ref{107}).
Substitution of representation (\ref{C2})-(\ref{C3}) into (\ref{C28}) yields:
\begin{eqnarray}\label{C30}
\mathcal{H}_{n,l}^{(4)}=\frac{N_{nl}^2\pi(\pi-2)}{3(2l-1)(2l+1)2^{l-1}}
\int_0^\pi (\sin \alpha)^{2l+2}\left[\cos\left(\frac{\alpha}{2}\right)+\sin\left(\frac{\alpha}{2}\right)\right]
C_{n/2-l}^{(l+1)}(\cos \alpha)\times
\nonumber~~~~~~~~~~\\
\left[
\frac{5}{(2l-3)\sin \alpha}F_{l,3}(\alpha)-\left(1-\frac{2}{\sin \alpha}\right)F_{l,1}(\alpha)-(2l-1)F_{l,-1}(\alpha)
\right]d\alpha.~~~~~~~
\end{eqnarray}
The orthogonality of the Legendre polynomials was used to derive the last formula.
Inserting expansion (\ref{C16}) into the lhs of Eq.(\ref{C1}), one obtains:
\begin{equation}\label{C32}
\left(\Lambda^2-32\right)\psi_{4,1}^{(2d)}=\sum_{nl}\mathcal{T}_{n,l}(n-4)(n+8)Y_{nl}(\alpha,\theta).~~~~~~~~~~~~~~~~~
\end{equation}
According to Eq.(\ref{C1}), the right-hand sides of Eq.(\ref{C32}) and Eq.(\ref{C25}) can be equated, what yields:
\begin{equation}\label{C33}
\mathcal{T}_{n,l}=\frac{\mathcal{H}_{n,l}^{(4)}}{(n-4)(n+8)}.~~~~~~~~~~~~~~~~~
\end{equation}
Thus, for the requested case of $n=2l$ ($l>2$), one obtains, using (\ref{C33}) and (\ref{C30}):
\begin{equation}\label{C35}
\mathcal{T}_{2l,l}=\frac{(\pi-2)2^{l-1}(l+1)(l!)^2}{3\pi^2(2l-1)(l-2)(l+4)(2l+1)!}\mathcal{P}_{3}(l),
\end{equation}
where
\begin{eqnarray}\label{C36}
\mathcal{P}_{3}(l)=
\int_0^\pi (\sin \alpha)^{2l+2}\left[\cos\left(\frac{\alpha}{2}\right)+\sin\left(\frac{\alpha}{2}\right)\right]\times
\nonumber~~~~~~~~~~~~~~~~~~~~~~~~~~~~~~~~~~\\
\left[
\frac{5}{(2l-3)\sin \alpha}F_{l,3}(\alpha)-\left(1-\frac{2}{\sin \alpha}\right)F_{l,1}(\alpha)-(2l-1)F_{l,-1}(\alpha)
\right]d\alpha.~~~
\end{eqnarray}
Integral (\ref{C36}) can be taken in the explicit form. The result is
\begin{eqnarray}\label{C37}
\mathcal{P}_{3}(l)=-\frac{2^{l+1}}{(l+3)(2l-3)(2l+3)(2l+5)}\times
\nonumber~~~~~~~~~~~~~~~~~~~~~~~~~~~~~~~~~~~~~~~~~~~~~~~~~~~~~~~~~~~~~~~~~~~~~~~~~~\\
\left\{
\frac{30}{l+1}-\frac{26}{l+2}+13-4l\left[47-2l(2l(l+3)-9)\right]+2^{l+1}\times
\right.
\nonumber~~~~~~~~~~~~~~~~~~~~~~~~~~~~~~~~~~~~~~~~~~~\\
\left.
\left[
(l+1)\left[4l(l(4l+3)-17)+45\right]B_{\frac{1}{2}}\left(l+\frac{3}{2},\frac{1}{2}\right)+
8l\left[l(l(4l(l+4)+3)-56)-62\right]B_{\frac{1}{2}}\left(l+\frac{3}{2},\frac{3}{2}\right)
\right]
\right\},
\nonumber\\
\end{eqnarray}
where $B_z(a,b)$ is the incomplete beta function.
Equating the right-hand sides of Eq.(\ref{C20}) and Eq.(\ref{C35}), one obtains the required coefficient in the form
\begin{equation}\label{C40}
A_2\equiv A_2(l)=\frac{1}{\mathcal{P}_{2}(l)}\left[\frac{(\pi-2)2^{-3(l+2)}}{3\pi(2l-1)(l-2)(l+4)}\mathcal{P}_{3}(l)-\mathcal{P}_{1}(l)\right].
\end{equation}
It may be verified that for \textbf{odd} values of $l$, the coefficients $A_2(l)$ equal zero.
For \textbf{even} values of $l$, formula (\ref{C40}) yields
\begin{equation}\label{C41}
A_2(l)=-\frac{(\pi-2)}{\pi^2}\mathcal{A}(l),
\end{equation}
where
\begin{eqnarray}\label{C42}
\mathcal{A}(4)=\frac{5515\pi-11648}{2948400},~\mathcal{A}(6)=\frac{191095\pi-396032}{1378377000},
~\mathcal{A}(8)=\frac{66779345\pi-137592832}{4085509428000},~~
\nonumber\\
\mathcal{A}(10)=\frac{59227659\pi-121716736}{25014766104900}.~~~~~~~~~~~~~~~~~~~~~~~~~~~~~~~~~~~~
\end{eqnarray}

\section{}\label{SD}

The IFRR (\ref{703}) for subcomponent $\psi_{3,0}^{(2c)}$ reads
\begin{equation}\label{D1}
\left(\Lambda^2-21\right)\psi_{3,0}^{(2c)}=h_{3,0}^{(2c)},
\end{equation}
where the rhs $h_{3,0}^{(2c)}$ is defined by Eq.(\ref{755}). To find a suitable solution in the form of expansion (\ref{726}), let us present the rhs of Eq.(\ref{D1}) in the form
\begin{equation}\label{D2}
h_{3,0}^{(2c)}\equiv -\frac{4\xi}{3 \sin \alpha}=\sum_{l=0}^\infty P_l(\cos \theta)(\sin \alpha)^l \textmd{h}_l(\alpha),
\end{equation}
where, using expansion (\ref{A7}) for $\nu=1$, one obtains
\begin{equation}\label{D3}
\textmd{h}_l(\alpha)=\frac{2^{2-l}}{3(2l-1)\sin \alpha}~_2F_1\left(\frac{l}{2}-\frac{1}{4},\frac{l}{2}+\frac{1}{4};l+\frac{3}{2};\sin^2\alpha\right).
\end{equation}
Inserting expansions (\ref{D2}) and (\ref{726}) into the IFFR (\ref{D1}), equating then the factors for the Legendre polynomials $P_l(\cos \theta)$ of the same order, and turning to the variable $\rho=\tan(\alpha/2)$, one obtains the inhomogeneous differential equation
\begin{equation}\label{D4}
\left(1+\rho^2\right)^2\phi_l''(\rho)+2\rho^{-1}\left[1+\rho^2+l(1-\rho^4)\right]\phi_l'(\rho)+(3-2l)(7+2l)\phi_l(\rho)=-h_l(\rho),
\end{equation}
which presents Eq.(\ref{335}) with $k=3$, and with changing $h(\rho)$ for $h_l(\rho)\equiv\textmd{h}_l(\alpha) $, and $g(\rho)$ for $\phi_l(\rho)$.
Simplification of the rhs of Eq.(\ref{D4}) yields for $0\leq\rho\leq1$:
\begin{equation}\label{D5}
h_l(\rho)= \frac{2^{1-l}\left(\rho^2+1\right)^{l+\frac{1}{2}}\left[(1-2l)\rho^2+2l+3\right]}{3(2l-1)(2l+3)\rho}.
\end{equation}
Simplifying the solutions (\ref{342}) of the homogeneous equation associated with Eq.(\ref{D4}), one obtains for $k=3$:
\begin{equation}\label{D6}
u_{3l}(\rho)=\frac{\left(\rho^2+1\right)^{l-\frac{3}{2}}}{\rho^{2l+1}}
\left[\frac{(2l+3)(2l+5)}{(2l-3)(2l-1)}\rho^4+\frac{2(2l+5)}{2l-1}\rho^2+1\right],
\end{equation}
\begin{equation}\label{D7}
v_{3l}(\rho)=\left(\rho^2+1\right)^{l-\frac{3}{2}}
\left[\frac{(2l-3)(2l-1)}{(2l+3)(2l+5)}\rho^4+\frac{2(2l-3)}{2l+3}\rho^2+1\right].
\end{equation}
The particular solution $\phi_l^{(p)}(\rho)$ of Eq.(\ref{D4}) can be obtained by formula (\ref{340}), but with changing definite integration for indefinite integration (that is putting the antiderivatives for the lower limits equal zero).
The result is presented by Eqs.(\ref{730})-(\ref{733}).
It is worth noting that the hypergeometric function included into Eq.(\ref{733}) can be presented in the form (see, 7.3.1.135 \cite{PRU3}):
\begin{equation}\label{D7a}
~_2F_1\left(1,l+1;l+2;-\rho^2\right)=-\frac{(l+1)}{\left(-\rho^2\right)^{l+1}}
\left[\ln\left(1+\rho^2\right)+\sum_{k=1}^l \frac{\left(-\rho^2\right)^{k}}{k}\right].
\end{equation}
It can be easily shown that the homogeneous solution $u_{3l}(\rho)$ is singular, whereas $v_{3l}(\rho)$ and the particular solution $\phi_l^{(p)}(\rho)$ are regular at the point $\rho=0~(\alpha=0)$ for any $l\geq0$.
The conclusion is:
the physical solution must be sought in the form (\ref{727}).
The problem is to find the coefficient $c_l$. To solve the problem we shall apply the coupling equation (\ref{546}), which for this case becomes:
\begin{equation}\label{D8}
\mathcal{O}_{2l,l}=\frac{(l+1)!2^{2(l+2)}}{\sqrt{\pi}\Gamma\left(l+\frac{3}{2}\right)}
\int_0^1\left[\phi_l^{(p)}(\rho)+c_l v_{3l}(\rho)\right]\frac{\rho^{2l+2}}{\left(\rho^2+1\right)^{2l+3}}d\rho,
\end{equation}
where $\mathcal{O}_{n,l}$ present the unnormalized HH expansion coefficients for subcomponent
\begin{equation}\label{D9}
\phi_{3,0}^{(2c)}(\alpha,\theta)=\sum_{nl}\mathcal{O}_{n,l}Y_{nl}(\alpha,\theta).
\end{equation}
For this case (odd value of $k=3$) the rhs of Eq.(\ref{D9}) presents the physical solution we are looking for (see, the end of Sec.\ref{S1}).
To derive the closed expression for $\mathcal{O}_{n,l}$, first, let us obtain the unnormalized HH expansion
\begin{equation}\label{D10}
h_{3,0}^{(2c)}(\alpha,\theta)=\sum_{nl}\mathcal{H}_{n,l}^{(3)}Y_{nl}(\alpha,\theta)
\end{equation}
for the rhs of Eq.(\ref{D1}), where by definition
\begin{equation}\label{D11}
\mathcal{H}_{n,l}^{(3)}=N_{nl}^2\int h_{3,0}^{(2c)}(\alpha,\theta)Y_{nl}(\alpha,\theta)d\Omega.
\end{equation}
Inserting expansion (\ref{D2})-(\ref{D3}) into (\ref{D11}), one obtains
\begin{equation}\label{D12}
\mathcal{H}_{n,l}^{(3)}=\frac{N_{nl}^2\pi^2 2^{3-l}}{3(2l-1)(2l+1)}
\int_0^\pi~_2F_1\left(\frac{l}{2}-\frac{1}{4},\frac{l}{2}+\frac{1}{4};l+\frac{3}{2};\sin^2\alpha\right)
\left(\sin \alpha\right)^{2l+1}C_{n/2-l}^{(l+1)}(\cos \alpha)d\alpha,~
\end{equation}
where the orthogonality of the Legendre polynomials was used. Substitution of expansion (\ref{D9}) into the lhs of Eq.(\ref{D1}) yields:
\begin{equation}\label{D13}
\left(\Lambda^2-21\right)\psi_{3,0}^{(2c)}=\sum_{nl}\mathcal{O}_{n,l}(n-3)(n+7)Y_{nl}(\alpha,\theta).~~~~~~~~~~~~~~~~~
\end{equation}
According to Eq.(\ref{D1}), the right-hand sides of Eq.(\ref{D10}) and Eq.(\ref{D13}) can be equated, what yields:
\begin{equation}\label{D14}
\mathcal{O}_{n,l}=\frac{\mathcal{H}_{n,l}^{(3)}}{(n-3)(n+7)}.~~~~~~~~~~~~~~~~~
\end{equation}
Thus, inserting (\ref{D12}) into Eq.(\ref{D14}) for $n=2l~(l\geq0)$, one obtains:
\begin{eqnarray}\label{D15}
\mathcal{O}_{2l,l}=
\frac{2^{l+4}l!(l+1)!}{3\pi(2l-3)(2l-1)(2l+7)(2l+1)!}\times
\nonumber~~~~~~~~~~~~~~~~~\\
\int_0^{\pi/2}~_2F_1\left(\frac{l}{2}-\frac{1}{4},\frac{l}{2}+\frac{1}{4};l+\frac{3}{2};
\sin^2\alpha\right)
\left(\sin \alpha\right)^{2l+1}d\alpha=
\nonumber~~~~~~~~~~~~~~~~~~~~~~~~~~~~~~~\\
\frac{2^{2-l}l!(l+1)!}{3(2l-3)(2l-1)(2l+7)\Gamma^2(l+3/2)}
~_3F_2\left(\frac{l}{2}-\frac{1}{4},\frac{l}{2}+\frac{1}{4},l+1;l+\frac{3}{2},l+\frac{3}{2};1\right).~~
\end{eqnarray}
Equating the right-hand sides of Eqs.(\ref{D8}) and (\ref{D15}), we find the required coefficient,
\begin{equation}\label{D16}
c_l=\frac{\mathcal{M}_1(l)-\mathcal{M}_2(l)}{\mathcal{M}_3(l)},
\end{equation}
where
\begin{equation}\label{D17}
\mathcal{M}_1(l)=\frac{2^{-3l-2}l!\sqrt{\pi}}{3(2l-3)(2l-1)(2l+7)\Gamma(l+3/2)}
~_3F_2\left(\frac{2l-1}{4},\frac{2l+1}{4},l+1;l+\frac{3}{2},l+\frac{3}{2};1\right),~~
\end{equation}
\begin{equation}\label{D18}
\mathcal{M}_3(l)\equiv\int_0^1v_{3l}(\rho)\frac{\rho^{2l+2}}{\left(\rho^2+1\right)^{2l+3}}d \rho=
\frac{2^{-l-\frac{3}{2}}(2l+1)}{(2l+3)(2l+7)}.~~~~~~~~~
\end{equation}
According to representation (\ref{730})-(\ref{733}), function $\mathcal{M}_2(l)$ can be presented in the form
\begin{eqnarray}\label{D19}
\mathcal{M}_2(l)\equiv \int_0^1\phi_{l}^{(p)}(\rho)\frac{\rho^{2l+2}}{\left(\rho^2+1\right)^{2l+3}}d \rho=
\nonumber~~~~~~~~~~~~~~~~~~~~~~~~~~~~~~~~~~~~~~~~~~~~~\\
\frac{2^{-l}}{3(2l-3)(2l-1)(2l+3)(2l+5)}
\left[2\mathcal{M}_{21}(l)+\frac{2\mathcal{M}_{22}(l)+\mathcal{M}_{23}(l)}{2l+1}\right],~~~~~~~
\end{eqnarray}
where
\begin{equation}\label{D20}
\mathcal{M}_{21}(l)=-\frac{1}{6}\left[
\frac{13-4l^2}{2^{l+\frac{5}{2}}}+\frac{8l^3+28l^2-2l-79}{l+2}~_2F_1\left(l+2,l+\frac{9}{2};l+3;-1\right)
\right],
\end{equation}
\begin{eqnarray}\label{D21}
\mathcal{M}_{22}(l)=\frac{(2l+1)(2l+5)}{(2l+3)(2l+7)^2}
\bigg \{ \frac{(2l+3)}{2^{l+\frac{7}{2}}}
\left[\pi(2l+7)-\frac{4(4l^2+24l+23)}{(2l+1)(2l+5)}\right]-
\nonumber~~~~~~~~~~~~~~~~\\
\frac{2\sqrt{\pi}(l+1)!}{\Gamma\left(l+\frac{1}{2}\right)}
+2^{\frac{3}{2}}(l+1)~_2F_1\left(-\frac{1}{2},-l;\frac{1}{2};\frac{1}{2}\right)
\bigg\},~~~~~~~~
\end{eqnarray}
\begin{equation}\label{D22}
\mathcal{M}_{23}(l)=(-1)^{l+1}\left[\mathcal{D}_l+\sum_{k=1}^l\frac{(-1)^k}{k}\mathcal{G}_{kl}\right].
~~~~~~~~~~~~~~~~~~~~~~~~~~~~~~~~~~~~~~~~
\end{equation}
In the last function we have introduced the following notations:
\begin{eqnarray}\label{D23}
\mathcal{D}_{l}=\frac{2^{-l-\frac{3}{2}}}{(2l+3)(2l+5)(2l+7)^2}
\bigg\{181\times 2^{l+\frac{5}{2}}-525\ln 2-1046+
l \Big[2^{l+\frac{11}{2}}(l+3)(11+2l(l+3))
\nonumber~~~~~\\
-8(195+4l(31+l(l+9)))-
2(985+4l(303+2l(83+l(21+2l))))\ln 2\Big]
\bigg \},~~~~~~~~~~
\end{eqnarray}
\begin{eqnarray}\label{D24}
\mathcal{G}_{kl}=\frac{(2l-3)(2l-1)}{2(k+1)}~_2F_1\left(k+1,l+\frac{9}{2};k+2;-1\right)+(2l+5)\times
\nonumber~~~~~~~~~~~~~~~~~~~~~~~~~~\\
\left[
\frac{2l-3}{k+2}~_2F_1\left(k+2,l+\frac{9}{2};k+3;-1\right)+
\frac{2l+3}{2(k+3)}~_2F_1\left(k+3,l+\frac{9}{2};k+4;-1\right)
\right].~~~
\end{eqnarray}
To derive formulas (\ref{D22})-(\ref{D24}), the representation (\ref{D7a}) was used.
\newpage

\begin{table}
\begin{center}
\caption{The simplest subcomponents of the solutions, and the corresponding rhs of the FRR.  }
\begin{tabular}{|c|c|c|c|c|}
\hline
{\normalsize $k$ }& {\normalsize $p$ }&{\normalsize  $j$} & { $h_{k,p}^{(j)}$} & { $\psi_{k,p}^{(j)}$}\tabularnewline
\hline
\hline
 $1$ & $0$ & $0$ &  $-2/\xi$ &  $\xi/2$ \tabularnewline
\hline
$3$ & $1$ &  $1$ &  $\frac{2(\pi-2)}{3\pi}\left(\xi^{-1}-\xi\right) $  & $\frac{(\pi-2)}{36\pi}\left(5\xi^2-6\right)\xi$ \tabularnewline
\hline
 $3$ &  $0$ &  $0$ & $E\xi+\frac{2E-1}{6}\xi^{-1}$ & $ \xi\left[\frac{1-2E}{24}+\frac{(E-2)}{72}\xi^2\right]$\tabularnewline
\hline
 $3$ &$0$ &  $1a$ & $-\frac{5(\pi-2)}{3\pi} \xi$  & $\frac{5(\pi-2)}{72\pi} \xi^3$ \tabularnewline
\hline
$3$ &$0$ &  $2a$ & $-\frac{2}{3} \left(2\xi+\xi^{-1}\right)$  & $\frac{1}{6}\xi\left(1-\frac{1}{3}\xi^2\right)$ \tabularnewline
\hline
$4$ &$1$ &  $1$ & $\frac{\pi-2}{18\pi}\left[6(1-2E)+(12E-5)\xi^2\right]$  & $\frac{(\pi-2)}{2880 \pi}
\left[3(32E-15)-8(12E-5)\xi^2\right]$ \tabularnewline
\hline
\hline
 $3$ & $0$ & $1b$ & $\frac{1}{3}\varsigma\left[\frac{1-2E}{\sin\alpha}+2(1-3E)\right] $ &  $\frac{1}{36}\varsigma\left[4E-1+(E-1)\sin \alpha\right]$ \tabularnewline
\hline
 $3$ & $0$ & $3$ & $\frac{2}{3}\left(\frac{2}{\sin \alpha}+3\right)\varsigma $ &  $-\frac{1}{36}\left(2+5\sin \alpha\right)\varsigma$ \tabularnewline
\hline
 $4$ & $1$ & $2b$ & $\frac{2(\pi-2)(5\pi-14)}{45\pi^2}\left(3-4\sin^2\alpha\right) $ &  $\frac{(\pi-2)(5\pi-14)}{540\pi^2}\csc\alpha\left[\alpha\cos(3\alpha)-\frac{1}{6}\sin(3\alpha)\right]$ \tabularnewline
\hline
 $4$ & $1$ & $3$ & $\frac{(\pi-2)}{3\pi}\left[2+\tan(\alpha/2)+\cot(\alpha/2)\right]\sin \alpha\cos \theta $ &
  $-\frac{\pi-2}{120\pi}\left(4+5\sin \alpha\right)\sin \alpha\cos \theta $ \tabularnewline
\hline
\end{tabular}
\label{T1}
\end{center}
\end{table}

\begin{table}
\caption{Numerical coefficients $\mu_{ln}^{(1)}$ (columns 2-6) and $\lambda_l$ included into Eq.(\ref{C14}) .}
\begin{tabular}{|c||c|c|c|c|c||c|}
\hline
{\footnotesize $l \diagdown n$} &{\footnotesize 0 } & {\footnotesize 1 } & {\footnotesize 2 } & {\footnotesize 3 } & {\footnotesize 4 } & {\footnotesize $\lambda_l$}\tabularnewline
\hline
\hline
{\footnotesize 3 } &{\footnotesize -522720 } &{\footnotesize -1051985 } &{\footnotesize 255552 } & {\footnotesize -90435 } &{\footnotesize 227872 } & {\footnotesize 495 } \tabularnewline
\hline
{\footnotesize 4 } &{\footnotesize -3953664 } &{\footnotesize -8398510 } &{\footnotesize 3727360 } & {\footnotesize -419475 } &{\footnotesize 3335168 } & {\footnotesize 2002 } \tabularnewline
\hline
{\footnotesize 5 } &{\footnotesize -30351360 } &{\footnotesize -97464675 } &{\footnotesize 44154880 } & {\footnotesize -6969375 } &{\footnotesize 39739392 } & {\footnotesize 6552 } \tabularnewline
\hline
{\footnotesize 6 } &{\footnotesize -2882764800 } &{\footnotesize -9137296644 } &{\footnotesize 5778063360 } & {\footnotesize -189629055 } &{\footnotesize 5234966528 } & {\footnotesize 235620 } \tabularnewline
\hline
{\footnotesize 7 } &{\footnotesize -270052392960 } &{\footnotesize -1196759060145 } &{\footnotesize 696562155520 } & {\footnotesize -65054175615 } &{\footnotesize 635252473856 } & {\footnotesize 7759752 } \tabularnewline
\hline
{\footnotesize 8 } &{\footnotesize -2853364039680 } &{\footnotesize -12357527011830 } &{\footnotesize 9051962343424 } & {\footnotesize -67609913085 } &{\footnotesize 8305992794112 } & {\footnotesize 27387360 } \tabularnewline
\hline
{\footnotesize 9 } &{\footnotesize -5615652962304 } &{\footnotesize -31944952238815 } &{\footnotesize 21219319480320 } & {\footnotesize -1350136582575 } &{\footnotesize 19579564720128 } & {\footnotesize 17341632 } \tabularnewline
\hline
{\footnotesize 10 } &{\footnotesize -132056403148800 } &{\footnotesize -733168078648720 } &{\footnotesize 580357875302400 } & {\footnotesize 1595198130669 } &{\footnotesize 538208290996224 } & {\footnotesize 127481640 } \tabularnewline
\hline
\end{tabular}
\label{T2}
\end{table}

\begin{table}
\caption{Numerical coefficients $\mu_{ln}^{(2)}$ and $\mu_{ln}^{(3)}$ included into Eq.(\ref{C14}) .}
\begin{tabular}{|c||c|c|}
\hline
{\footnotesize $l \diagdown \mu$} & {\footnotesize $\mu_{nl}^{(2)} $}&{\footnotesize $\mu_{nl}^{(3)} $} \\
\hline
\hline
{\footnotesize 3} & {\footnotesize $\{-72765,-412335,-960498,-2625678\}$}&{\footnotesize $\{1\}$ } \\
\hline
{\footnotesize 4} & {\footnotesize $\{1273965,5945170,9597203,3639900,-14747117\}$}&{\footnotesize $\{1,-1\}$ } \\
\hline
{\footnotesize 5} & {\footnotesize $\{-8513505,-34999965,-45846801,-10135125,10135125,-146271983\}$}&{\footnotesize $\{1,-\frac{14}{9},1\}$ } \\
\hline
\multirow{2}{*}{\footnotesize 6 }
 &{\footnotesize $\{1456717185,5473725180,6250641012,983757060,-777756650,$} &{\footnotesize $\{1,-\frac{21}{11},\frac{21}{11},-1\}$ } \\
 &{\footnotesize $983757060,-13174180236\}$}&\\
  \hline
\multirow{2}{*}{\footnotesize 7 }
 &{\footnotesize $\{-101948591745,-358127104335,-370912265724,-46034728740,31617756170,$} &{\footnotesize $\{1,-\frac{28}{13},\frac{378}{143},-\frac{28}{13},$ } \\
 &{\footnotesize $-31617756170,46034728740,-1585784814084\}$}&{\footnotesize $1\}$}\\
  \hline
\multirow{2}{*}{\footnotesize 8 }
 &{\footnotesize $\{1804845357315,6016151191050,5790931172103,602937350136,-374632125450,$} &{\footnotesize $\{1,-\frac{7}{3},\frac{42}{13},-\frac{42}{13},$ } \\
 &{\footnotesize $328821494780,-374632125450,602937350136,-16476813837561\}$}&{\footnotesize $\frac{7}{3},-1\}$}\\
  \hline
\multirow{2}{*}{\footnotesize 9 }
 &{\footnotesize $\{-2508790869585,-8018292387105,-7290251115147,-663280843995,381499104294,$} &{\footnotesize $\{1,-\frac{42}{17},\frac{63}{17},-\frac{924}{221},$ } \\
 &{\footnotesize $-306553004730,306553004730,-381499104294,663280843995,-39837938484597\}$}&{\footnotesize $\frac{63}{17},-\frac{42}{17},1\}$}\\
  \hline
\multirow{3}{*}{\footnotesize 10 }
 &{\footnotesize $\{95748321954285,295643941472880,256791268415682,21004496987760,$} &{\footnotesize $\{1,-\frac{49}{19},\frac{1323}{323},-\frac{1617}{323},$ } \\
 &{\footnotesize $-11353909326045,8538680495904,-7845295711140,8538680495904,$}&{\footnotesize $\frac{1617}{323},-\frac{1323}{323},\frac{49}{19},-1\}$}\\
 &{\footnotesize $-11353909326045,21004496987760,-931653655078718\}$}&\\
  \hline
\end{tabular}
\label{T3}
\end{table}


\begin{thebibliography}{99}

\bibitem{B35} J. H. Bartlett, J. J. Gibbons and C. G. Dunn, "The Normal Helium Atom", Phys. Rev. \textbf{47}, 679-680 (1935).
\bibitem{B37} J. H. Bartlett, "The Helium Wave Equation", Phys. Rev. \textbf{51}, 661-669 (1937).

\bibitem{FOCK} V. A. Fock, Izv. Akad. Nauk SSSR, Ser. Fiz. \textbf{18}, 161 (1954).

\bibitem{ES1} A. M. Ermolaev and G. B. Sochilin, "The ground state of two-electron atoms and ions",
Sov. Phys. - Doklady \textbf{9}, 292-295 (1964).

\bibitem{DEM} Y. N. Demkov and A. M. Ermolaev, "Fock expansion for the wave functions of a system of charged particles",
 Sov. Phys. JETP \textbf{9}, 633-635 (1959).

\bibitem{TUL} A. V. Tulub, "Application of the Fock expansion to the theory of van der Waals forces" , Sov. Phys.-Dokl. \textbf{13}, 936-938 (1969).

\bibitem{ES2} A. M. Ermolaev, Vest. Len. Univ. \textbf{16}, 19-33 (1961).


\bibitem{DAV} C. W. David, "$\Psi_{20}$ in the Fock expansion for He $^1S$ state wavefunctions", J. Chem. Phys. \textbf{63}, 2041-2044 (1975).

\bibitem{PLUV} Ph. Pluvinage, "Premiers termes du d\'{e}veloppement de Fock pour les \'{e}tats $S$ de $HeI$
et de sa s\'{e}quence iso\'{e}lectronique", J. Physique \textbf{43}, 439-458 (1982).

\bibitem{FMS} J. M. Feagin, J. Macek, and F. Starace, "Use of the Fock expansion for $^1S$-state wave functions of the two-electron atoms ans ions", Phys. Rev. A \textbf{32}, 3219-3230 (1985).

\bibitem{AM1} P. C. Abbott and E. N. Maslen, "Coordinate systems and analytic expansions for three-body atomic wavefunctions: I. Partial summation for the Fock expansion in hyperspherical coordinates", J. Phys. A: Math. Gen. \textbf{20}, 2043-2075 (1987).

\bibitem{GAM2} J. E. Gottschalk, P. C. Abbott and E. N. Maslen, "Coordinate systems and analytic expansions for three-body atomic wavefunctions: II. Closed form wavefunction to second order in $r$", J. Phys. A: Math. Gen. \textbf{20}, 2077-2104 (1987).

\bibitem{GM3} J. E. Gottschalk and E. N. Maslen, "Coordinate systems and analytic expansions for three-body atomic wavefunctions: III. Derivative continuity via solution to Laplace's equation", J. Phys. A: Math. Gen. \textbf{20}, 2781-2803 (1987).

\bibitem{ATH} P. C. Abbott, "Analytic solutions to the few-body Schr\"{o}dinger equation using hyperspherical coordinates", Thesis (1986).

\bibitem{FOR} R. C. Forrey, "Compact representation of helium wave functions in perimetric and hyperspherical coordinates", Phys. Rev. A \textbf{69}, 022504 (2004).

\bibitem{LEZ1} E. Z. Liverts, "Two-particle atomic coalescences", Phys. Rev. A \textbf{89}, 032506 (2014).

\bibitem{MYE} C. R. Myers, C. J. Umrigar, J. P. Sethna, J. D. Morgan III, "Fock's expansion, Kato's cusp conditions, and the exponential ansatz", Phys. Rev. A \textbf{44}, 5537-5546 (1991).

\bibitem{LL1} J. Cho, "Lobachevsky function and dilogarithm function", Lecture Note, 2007.

\bibitem{LL2} J. G. Racliffe, "Foundations of hyperbolic manifolds", (2nd ed.). "Graduate Texts in Mathematics", 149, Springer, 2006.

\bibitem{Sack} R. A. Sack, "Generalization of Laplace's expansion to arbitrary powers and functions of the distance between two points", J. Math. Phys. \textbf{5}, 245-251 (1964).

\bibitem{PRU} A. P. Prudnikov, Yu. A. Brychkov and O. I. Marichev, " Integrals and Series. Vol 1. Elementary Functions", Gordon and Breach S. P., New-York, 1986.

\bibitem{PRU3} A. P. Prudnikov, Yu. A. Brychkov and O. I. Marichev, " Integrals and Series. Vol 3. More Special Functions", Gordon and Breach S. P., New-York, 1986.

\bibitem{RED} E. D. Rainville, "Special Functions, New-York, 1960.


\newpage
\end{thebibliography}
\end{document}